\documentclass[a4paper,11pt]{article}
\pdfoutput=1 
\usepackage{jcappub}
\usepackage{mathtools}
\usepackage{graphicx}
\usepackage{natbib}
\usepackage{multirow}
\usepackage{amsmath}
\usepackage[amssymb]{SIunits}
\usepackage{empheq}
\usepackage{booktabs}

\usepackage{color}

\newcommand{\Omegadm}{\Omega_{\textrm{dm}}}
\newcommand{\omegadm}{\omega_{\textrm{dm}}}
\newcommand{\Omegam}{\Omega_{\textrm{m}}}
\newcommand{\omegam}{\omega_{\textrm{m}}}

\newcommand{\rhodcdm}{\rho_{\rm dcdm}}
\newcommand{\rhodr}{\rho_{\rm dr}}
\newcommand{\Omegadcdm}{\Omega_{\rm dcdm}}
\newcommand{\Omegadr}{\Omega_{\rm dr}}
\newcommand{\Omegasdm}{\Omega_{\rm sdm}}
\newcommand{\omegadcdm}{\omega_{\rm dcdm}}

\newcommand{\omegasdm}{\omega_{\rm sdm}}
\newcommand{\deltadcdm}{\delta_{\rm dcdm}}
\newcommand{\thetadcdm}{\theta_{\rm dcdm}}
\newcommand{\deltadr}{\delta_{\rm dr}}
\newcommand{\thetadr}{\theta_{\rm dr}}
\newcommand{\mpsi}{\mathfrak{m}_{\psi}}
\newcommand{\mcont}{\mathfrak{m}_{\rm cont}}
\newcommand{\mshear}{\mathfrak{m}_{\rm shear}}

\newcommand{\fdcdm}{f_{\rm dcdm}}
\newcommand{\Gammadcdm}{\Gamma_{\rm dcdm}}

\usepackage[T1]{fontenc} 

\title{\boldmath A fresh look at linear cosmological constraints on a decaying dark matter component}

\author[a,b]{Vivian Poulin,}
\author[a,b]{Pasquale D. Serpico,}
\author[b]{Julien Lesgourgues}

\affiliation[a]{LAPTh, Universit\'e Savoie Mont Blanc \& CNRS, BP 110,\\ F-74941 Annecy-le-Vieux Cedex, France.}
\affiliation[b]{Institute for Theoretical Particle Physics and Cosmology (TTK), \\ RWTH Aachen University, D-52056 Aachen, Germany.}

\emailAdd{Vivian.Poulin@lapth.cnrs.fr}
\emailAdd{Pasquale.Serpico@lapth.cnrs.fr}
\emailAdd{Julien.Lesgourgues@physik.rwth-aachen.de}

\abstract{We consider a cosmological model in which a fraction $\fdcdm$ of the Dark Matter (DM) is allowed to decay in an invisible relativistic component, and compute the resulting constraints on both the decay width (or inverse lifetime) $\Gammadcdm$ and $\fdcdm$ from purely gravitational arguments. We report a full derivation of the Boltzmann hierarchy, correcting a  mistake in previous literature, and compute the impact of the decay---as a function of the lifetime---on the CMB and matter power spectra. From CMB only, we obtain that no more than 3.8\% of the DM could have decayed in the time between recombination and today (all bounds quoted at 95\% CL). We also comment on the important application of this bound to the case
where primordial black holes constitute DM, a scenario notoriously difficult to constrain. For lifetimes longer than the age of the Universe, the bounds can be cast as $\fdcdm\Gammadcdm < 6.3\times10^{-3}$ Gyr$^{-1}$. For the first time, we also checked that degeneracies with massive neutrinos are broken when information from the large scale structure is used. Even secondary effects like CMB lensing suffice to this purpose. 
Decaying DM models have been invoked to solve a possible tension between low redshift astronomical measurements of $\sigma_8$ and $\Omega_{\rm m}$ and the ones inferred by Planck. We reassess this claim finding that with the most recent BAO, HST and $\sigma_8$ data extracted from the CFHT survey, the tension is only slightly reduced despite the two additional free parameters. Nonetheless, the existing tension explains why the bound on $\fdcdm\Gammadcdm$ loosens to $\fdcdm\Gammadcdm$ < 15.9$\times10^{-3}$ Gyr$^{-1}$ when including such additional data. The bound however improves to $\fdcdm\Gammadcdm$ < 5.9$\times10^{-3}$ Gyr$^{-1}$ if only data consistent with the CMB are included. This highlights the importance of establishing whether the tension is due to real physical effects or unaccounted systematics, for settling the reach of achievable constraints on decaying DM.}

\begin{document}
\hfill{\small LAPTH-027/16}\\
\maketitle
\flushbottom
\section{Introduction}
\label{sec:intro}

The current concordance model of cosmology  ($\Lambda$CDM, supplemented by the inflationary paradigm) has been established very robustly over the past couple of decades, surviving a large number of tests and cross-checks. Nonetheless, it remains a {\it parametric} model, with most of the energy content of the universe in a ``dark'' sector whose nature
remains puzzling. In particular, although there is plenty of proposed candidates for what makes the dark matter (DM) of the Universe, none of them could  be detected through smoking gun probes independent from the gravitational ones, and thus DM lacks identification.

In the quest for the nature of DM, cosmology itself provides useful diagnostics, being sensitive to a large range of spatial and time scales.
Some of us have recently revisited~\cite{Audren14} the (mostly) linear cosmological signatures of a finite DM lifetime, purely via gravitational effects. Often, cosmological constraints are
 not as constraining as observables targeted to specific DM models (such as gamma-ray fluxes for vanilla WIMP DM models). Yet, they are valuable as
essentially based on linear theory, thus setting the robust and model-independent yardstick against which gauging the strength of other constraints---in general depending on non-linear physics, non-gravitational interactions, and a number of astrophysical assumptions.

In this article, we provide several improvements as well as an important generalisation over the treatment in~\cite{Audren14}. First and foremost, we allow for a fraction $\fdcdm$ of decaying DM smaller than unity. One immediate consequence is that a much larger parameter space for the  decay width (or inverse lifetime) $\Gammadcdm$ is now open, with
a richer set of consequences on the CMB anisotropy pattern and power spectrum modifications which we shall duly describe.
One (perhaps phenomenologically compelling) motivation for such a refined study is the recurring recent claim that some tensions in global fits of cosmological observables, like CMB ones vs. the low-redshift determination of $\sigma_8$ and the Hubble parameter~\cite{Riess:2011yx,Heymans:2013fya,Ade:2015fva,Riess:2016jrr}, may be resolved due to a non-trivial time evolution of the DM content of the universe, such as the one associated with a (partial) decay of the DM constituents~\cite{Enqvist:2015ara,Berezhiani:2015yta,Chudaykin:2016yfk}.
From a theoretical point of view, the  case $\fdcdm<1$ is also interesting, with several possible physical interpretations:
1) DM may be multi-component, with one unstable component disintegrating into (dark) radiation;
2) DM may decay into several particles, including  a fraction $(1-\fdcdm)$ of cold daughter particles and a fraction $\fdcdm$ of (dark) relativistic daughter particles, such as neutrinos, gravitons, or some BSM (beyond the standard model) species.

Throughout the paper, we shall assume that the stable fraction of the DM (or byproduct of the DM disintegration)  is exactly cold. This can be certainly achieved in scenario 1), while being an approximation in scenario 2):  the recoil received by the daughter non-relativistic particle(s) is responsible for some velocity dispersion in the daughter DM phase-space distribution.  A number of specific implications of these so-called superWIMP candidates is known since more than a decade~\cite{Cembranos:2005us}. Dealing with this scenario would require making specific assumption on the final state kinematics. Qualitatively, we expect our bounds to be still valid for scenarios of type 2) as well, since  the sizable DM free streaming would impose further and typically more stringent constraints. The analysis presented in~\cite{Aoyama:2014tga}, which focuses on a scenario of type 2) (i.e. with a specific choice of DM two-body decay into radiation plus one massive relic) obtains indeed tighter constraints than the present work.

The generality of purely gravitational constraints does not imply, however, that such bounds are always ``weak'': for several particle physics candidates without sizable signals in non-gravitational channels, they are the strongest available ones. For instance, this is the case for
the majoron, a pseudo-goldstone boson associated to the breaking of the global lepton number symmetry, acquiring a small mass plausibly due to quantum gravity effects, whose
cosmological interest in a modern context has been revisited in~\cite{Lattanzi:2007ux} and references therein. Even within SUSY scenarios, if the lightest SUSY particle and
the next-to-lightest one (NLSP) are respectively gravitinos and right-handed sneutrinos (see e.g.~\cite{Allahverdi:2014bva}), or multiple sneutrino states (for a recent example see~\cite{Banerjee:2016uyt}), such bounds may be relevant. In SUSY scenarios incorporating
the Peccei-Quinn mechanism~\cite{Peccei:1977hh,Peccei:1977ur} to solve the strong CP problem, the role of dark radiation and unstable DM parents/byproducts might be played by
the axion~\cite{Weinberg:1977ma,Wilczek:1977pj}, its supersymmetric partner the axino, and gravitinos (see for instance~\cite{Choi:1996vz} for an early proposal in this sense, or~\cite{Baer:2014eja} for a modern review).
Note that our results may be of interest also to other scenarios, remarkably including also cases where DM {\it is not} unstable. For instance, whenever a
sizable or dominant fraction of the DM is in the form of primordial black holes, a non-trivial merger history may alter their initial mass function and the associated phenomenological
constraints. Yet, in the merger process a non-negligible fraction of their mass converts into gravitational waves, a form of dark radiation subject to the constraint discussed in this paper.

This article is structured as follows: in Sec.~\ref{sec:Equations}, for the sake of completeness and in order to correct an error present in past literature, we explicitly report the key equations solved. In Sec.~\ref{sec:effects} we describe the effects of the decaying DM models on CMB and power spectrum observables. We also discuss some potential degeneracies with other physical effects, notably the one of massive neutrinos. In Sec.~\ref{sec:results} we present our results, while in section~\ref{sec:conclu}, after a discussion, we report our conclusions.
\section{Boltzmann equations for the decaying Dark Matter}
\label{sec:Equations}
We wish here to recall the main equations describing the gravitational impact of the DM decay, assuming that the decay products are ultrarelativistic and invisible---hence dubbed ``dark radiation'', denoted with the subscript ``dr''.
Following the standard procedure, we consider small perturbations over a homogeneous background and hence, we split up the evolution equations for the energy density and momenta of the decaying cold DM (denoted with the subscript ``dcdm'') and its daughter radiation between zeroth and first order contributions. Higher orders terms are neglected. The overall DM
abundance is denoted with the subscript ``dm''; $\fdcdm$ is the ratio of the decaying DM fraction to the total one. Its complement to one is also dubbed stable DM fraction, denoted with subscript ``sdm''.

\subsection{Background equations}
To take dark matter decay into account, one can for instance modify the stress energy tensor of cold DM and dark radiation by respectively subtracting and adding a decay term.
By considering  the covariant conservation of $T_{\mu\nu}$ that follows from Bianchi identities, one would arrive at \cite{Kang93} (see also~\cite{Audren14})
\begin{eqnarray}
\rhodcdm '=-3\frac{a'}{a}\rhodcdm-a\Gammadcdm\rhodcdm~,\label{eq:rhodcdm}\\
\rhodr '= -4\frac{a'}{a}\rhodr + a\Gammadcdm\rhodcdm~.\label{eq:rhodr}
\end{eqnarray}
Above and henceforth, prime quantities denote a derivative with respect to conformal time; $\Gammadcdm$ is the decay rate defined with respect to proper time,  which in a specific model can be computed as customary by integrating over the phase space (the modulus square of) the transition matrix element.
Using the public version of the {\sc class}\footnote{\tt class-code.net}~\cite{Lesgourgues:2011re,Blas:2011rf} Einstein-Boltzmann solver, it is possible to specify the total fractional energy density in both dcdm and dr, either {\em today} ($\Omegadcdm +\Omegadr$) or {\em initially} ($(\Omegadcdm+\Omegadr)^{\rm ini}$). The latter parameter is defined precisely in~\cite{Audren14}, but can be loosely understood as  ``the value of the initial densities $\rho_{\rm dcdm}(t_{\rm ini})$ and $\rho_{\rm dr}(t_{\rm ini})$ such that if we were to take $\Gammadcdm=0$, we would get a fractional density $\Omegadcdm +\Omegadr=(\Omegadcdm+\Omegadr)^{\rm ini}$ today''; the splitting between  $\rho_{\rm dcdm}(t_{\rm ini})$ and $\rho_{\rm dr}(t_{\rm ini})$ is computed automatically in order to take consistently into account the tiny amount of transfer of energy from dcdm to dr between $t \longrightarrow 0$ and $t_{\rm ini}$, assuming that there is no dark radiation for $t \longrightarrow 0$. This parametrisation has the advantage of preserving the early cosmological evolution until DM starts to decay.

\subsection{Perturbation equations in gauge invariant variables}
In ref.~\cite{Audren14}, the scalar perturbations equations at zeroth and first order had been obtained starting from the continuity and Euler equations, which describes the exchange of energy and momenta between the decaying  DM and the dark radiation.
Here, we report the result starting  from the Boltzmann equation describing the evolution of the two species, which is  also  needed  to derive the Boltzmann hierarchy of the dark radiation.
The full Boltzmann equation, with decay term, written in terms of conformal time and momentum is
\begin{equation}\label{eq:Boltzmann1}
\frac{df}{d\tau}=\frac{\partial f}{\partial \tau}+\frac{\partial f}{\partial x^i}\frac{d x^i}{d\tau}+\frac{\partial f}{\partial q}\frac{dq}{d\tau}+\frac{\partial f}{\partial n^i}\frac{dn^i}{dt}=\pm a \Gammadcdm f_{\rm dcdm}\equiv \pm D\,,
\end{equation}
where the $-~(+)$ sign refers to the decaying  DM (dark radiation).
Let us stress an important point, that has been overlooked in a previous paper deriving bounds on the model we are dealing with \cite{Ichiki04}.  As long as one does not consider perturbations in the distribution, the previous form of the decay term still holds. However, the decay term $D$ is {\em not} a gauge invariant quantity. At the level of perturbations, such a simple form of the decay rate is only valid in the gauge {\em comoving} with the decaying DM, which is a restriction of the synchronous gauge to a system in which the velocity divergence of the decaying DM vanishes.
Note that, as pointed out in Ref.~\cite{Audren14}, the comoving gauge of the DM and decaying DM are the same for adiabatic initial conditions.
Hence we can work with the synchronous gauge comoving with all DM, in which metric perturbations read
\begin{equation}\label{eq:SyncGaugev2}
ds^2=a^2(\tau)\bigg\{-d\tau^2+(\delta_{ij}+H_{ij})dx^idx^j\bigg\}
\end{equation}
(with in Fourier space $H_{ij} = \hat{k}_i\hat{k}_jh+(\hat{k}_i\hat{k}_j-\frac{1}{3}\delta_{ij})6\eta$ ), and with the
newtonian gauge with metric perturbations
\begin{equation}\label{eq:NewtGaugev2}
ds^2=a^2(\tau)\bigg\{-(1+2\psi)d\tau^2+(1-2\phi)dx^idx^j\bigg\}\,.
\end{equation}
One can relate both gauges via the following identities
\begin{eqnarray}\label{eq:GaugeTransfo1}
\psi & =  \mathcal{H}\alpha+\alpha'\\
\phi & =  \eta-\mathcal{H}\alpha\\
\delta(new)& =  \delta(syn)  +\frac{\bar{\rho}'}{\bar{\rho}}\alpha\\
\theta(new) & = \theta(syn) +k^2\alpha\label{eq:GaugeTransfo2}
\end{eqnarray}
where $\alpha\equiv\frac{(6\eta+h)'}{2k^2}$. All definitions of the potentials and perturbations variables can be found in Ref.~\cite{Ma:1995ey}.
We will use these relations later in order to get the evolution equations in the newtonian gauge.
Starting with the decaying DM, in the comoving synchronous gauge one can re-express eq.~(\ref{eq:Boltzmann1}) in the following form:
\begin{equation}\label{BoltzmannDcdm}
\frac{\partial f_{\rm dcdm}}{\partial \tau}+\frac{\partial f_{\rm dcdm}}{\partial x^i}\frac{p^i}{E}+p\frac{\partial f_{\rm dcdm}}{\partial p}\bigg[\eta'-\frac{1}{2}\big(h'+6\eta'\big)\big(\hat{k}\cdot n\big)^2-\mathcal{H}\bigg]=- D
\end{equation}
where we have now traded $q$ for $p$ (the physical momentum) to let the expansion term appear explicitly.
Integrating the distribution over the phase-space and using the fact that decaying DM is pressureless, it is straightforward to get
\begin{eqnarray}\label{eq:BoltzmannDensityDcdm}
\rho_{\rm  dcdm}'+\frac{\partial (\rho_{\rm  dcdm} v_{\rm  dcdm}^i)}{\partial x^i} +\frac{1}{2}h'\rho_{\rm  dcdm}+3\mathcal{H}\rho_{\rm  dcdm}
& = & -a\Gamma \rho_{\rm  dcdm}\,.
\end{eqnarray}
We introduce the usual notation $\rho_{\rm  dcdm}=\bar{\rho}_{\rm  dcdm}[1+\delta_{\rm  dcdm}]$ and set in our gauge $\theta_{\rm  dcdm} = \partial_i v^i_{\rm  dcdm} = 0$.
At this level, if we were to collect all zeroth order terms, we would arrive at Eq.~(\ref{eq:rhodcdm}).
Collecting instead all first order terms and dividing by $\bar{\rho}_{\rm  dcdm}$, one gets
\begin{equation}\label{eq:ContinuityDecay}
\delta_{\rm  dcdm}'=-\frac{h'}{2}\,.
\end{equation}
In a similar way, taking the divergence of the first moment of eq.~(\ref{eq:BoltzmannDensityDcdm}) leads to
\begin{eqnarray}
\theta_{\rm dcdm}' & = & -\mathcal{H}\theta_{\rm dcdm},
\end{eqnarray}
which is consistent with keeping $\theta_{\rm dcdm}=0$ in the comoving synchronous gauge.
We can make use of relations (\ref{eq:GaugeTransfo1}-\ref{eq:GaugeTransfo2}) to express previous equations in the newtonian gauge:
\begin{eqnarray}
\delta_{\rm dcdm}^{(n)'}&=&-a\Gamma\phi+3\psi-\theta_{\rm dcdm}^{(n)}\,,\\
\theta_{\rm dcdm}^{(n)'}&=&-\mathcal{H}\theta_{\rm dcdm}^{(n)}+k^2\phi\,.
\end{eqnarray}
Following the development of Ref.~\cite{Audren14}, we introduce the gauge invariant variables $\mcont$ and $\mpsi$ to write these equations as
\begin{eqnarray}
\deltadcdm'&=&-\thetadcdm-\mathfrak{m}_{\rm{cont}}-a\Gamma\mathfrak{m}_{\psi}~,\\
\thetadcdm'&=&-\mathcal{H}\thetadcdm+k^2\mathfrak{m}_{\psi}~
\end{eqnarray}
where the expression of the metric source terms in the synchronous and newtonian gauges are given in table \ref{table:metricSourceTerms}.
\begin{table}[!h]
\centering
\begin{tabular}{l|cc}
	& Synchronous \quad& Newtonian\\
	\hline
   $\mathfrak{m}_{\rm{cont}}$  & $h'/2$ & $-3\phi'$ \\
   $\mathfrak{m}_{\psi}$ & 0& $\psi$ \\
   $\mshear$  & $(h'+6\eta')/2$ & 0 \\
 \end{tabular}
  \caption{\label{table:metricSourceTerms} Continuity and euler type of metric source terms for scalar perturbations in synchronous and Newtonian gauge.}
 \end{table}

Although these equations are sufficient to describe the dynamics of the decaying DM, one needs to write the full Boltzmann hierarchy to follow the perturbations in the dark radiation.
For that purpose, several strategies can be adopted; we follow again the formalism of Ref.~\cite{Audren14}. We introduce the perturbations of the integrated phase-space distribution function and expand it over Legendre polynomials $\mathcal{P}_\ell$ in the following way
\begin{equation}\label{eq:Fdr}
F_{\rm dr}\equiv\frac{\int dqq^3f_{\rm dr}^{(0)}\Psi_{\rm dr}}{\int dqq^3f_{dr}^{(0)}}r_{\rm dr}\equiv\sum_\ell(-i)^\ell(2\ell+1)F_{{\rm dr},\ell}(t,\vec{k})\mathcal{P}_\ell(\mu)
\end{equation}
where $\Psi_{\rm dr}$ is defined at the level of the perturbed phase-space distribution:
\begin{eqnarray}\label{eq:fPerturbedv2}
f_{\rm dr}(\vec{x},p,\vec{n},\tau)=f_{\rm dr}^{(0)}(p,\tau)(1+\Psi_{\rm dr}(\vec{x},p,\vec{n},\tau))\,,\nonumber\\
\end{eqnarray} and $r_{\rm dr}$ is defined as
\begin{equation}
r_{\rm dr}\equiv\frac{\bar{\rho}_{\rm dr}a^4}{\rho_{cr,0}}\,.
\end{equation}
$\rho_{cr,0}$ is the critical energy density today (where $a_0=1$), introduced to make $r_{\rm dr}$ dimensionless.
The use of $r_{\rm dr}$ will help us to cancel the time-dependence of $F_{\rm dr}$ due to the background distribution function $f_{dr}^{0}$.
The derivative of $r_{\rm dr}$ is simply
\begin{eqnarray}
r_{\rm dr}' & = & a\Gammadcdm \frac{\bar{\rho}_{\rm dcdm}}{\bar{\rho}_{\rm dr}}r_{\rm dr}\,,
\end{eqnarray}
(here we corrected a typo with respect to eq.~(2.16) Ref.~\cite{Audren14}).
Splitting between $m= 0,1,2$ and $m>2$ leads to the following gauge-independent hierarchy
\begin{eqnarray}\label{eq:HierarchyWithDecayBothGauges}
F_{\rm dr,0}' & = & -kF_{{\rm dr},1}-\frac{4}{3}r_{\rm dr}\mcont+r_{\rm dr}'(\delta_{\rm dcdm}+\mpsi)~,\\
F_{\rm dr,1}' & = & \frac{k}{3}F_{{\rm dr},0}-\frac{2k}{3}F_{{\rm dr},2}+\frac{4k}{3}r_{\rm dr}\mpsi+\frac{r_{\rm dr}'}{k}\theta_{\rm dcdm}~,\\
F_{\rm dr,2}' & = & \frac{2k}{5}F_{{\rm dr},1}-\frac{3k}{5}F_{{\rm dr},3}+\frac{8}{15}r_{\rm dr}\mshear~,\\
F_{\rm dr,\ell}' & = & \frac{k}{2\ell+1}\big(\ell F_{{\rm dr},\ell-1}-(\ell+1)F_{{\rm dr},\ell+1}\big)\qquad \ell>2.
\end{eqnarray}
The expression for $\mshear$ in newtonian and synchronous gauge is given in table \ref{table:metricSourceTerms}.
This set of equations must be truncated at some maximum multipole order $\ell_{\rm{max}}$. To do so, we use the improved truncation scheme of Ref.~\cite{Ma:1995ey} which has been generalised to spatial curvature in Ref.~\cite{Lesgourgues13}.\\
To check how the multipole moments transform, one can use their relations to the standard variables $\delta$ and $\theta$, as well as the gauge-invariant variable $\sigma$, which describes the anisotropic stress developing in the fluid:
\begin{eqnarray}\label{eq:FtoStandardVariables}
F_{{\rm dr},0} = r_{{\rm dr}}\delta_{{\rm dr}}~,\qquad\qquad F_{{\rm dr},1} = \frac{4r_{{\rm dr}}}{3k}\theta_{{\rm dr}}~,\qquad\qquad  F_{{\rm dr},2}=2\sigma r_{{\rm dr}}~.
\end{eqnarray}
Eqs.~(\ref{eq:GaugeTransfo1}-\ref{eq:GaugeTransfo2}) then immediately tell us how these moments change under a transformation from the synchronous to newtonian gauge:
\begin{eqnarray}
F_{{\rm dr},0}^{(s)}=F_{{\rm dr},0}^{(n)}-r_{{\rm dr}}\frac{\bar{\rho}_{{\rm dr}}'}{\bar{\rho}_{{\rm dr}}}\alpha~,\qquad\qquad F_{dr,1}^{(s)}  = F_{{\rm dr},1}^{(n)}-\frac{4r_{{\rm dr}}k}{3}\alpha~,
\end{eqnarray}
whereas $F_{{\rm dr},2}$ is gauge invariant.
Using eq.~(\ref{eq:FtoStandardVariables}), one can easily check that we get the same equations as the ones coming from considering energy and momentum conservation:
\begin{eqnarray}
\deltadr'&=&-\frac{4}{3}(\thetadr+\mcont)+a\Gammadcdm\frac{\rhodcdm}{\rhodr}(\deltadcdm-\deltadr+\mpsi)~,\\
\thetadr'&=&\frac{k^2}{4}\deltadr-k^2\sigma_{\rm dr}+k^2\mpsi-a\Gammadcdm\frac{3\rhodcdm}{4\rhodr}\bigg(\frac{4}{3}\thetadr-\thetadcdm\bigg)~.\\
\end{eqnarray}
Finally, one can compare our final set of equations in the newtonian gauge with the one of Ref.~\cite{Ichiki04}. It turns out that this reference omitted the $\mathfrak{m}_{\psi}$ term in the evolution equations of $\deltadcdm$ and $\deltadr$. Hence, even besides the fact that we are using more recent data, we do not expect our results to match the ones reported in~\cite{Ichiki04}.

\section{Cosmological effects of a decaying Dark Matter fraction}\label{sec:effects}
In this section we show the impact of the decaying DM on the CMB and matter power spectra from purely gravitational effects, as a function of the DM lifetime.
This was already described in Ref.~\cite{Audren14} in the case of long-lived fully-decaying DM, but we aim at generalizing this study to the case of multi-component DM splitted as $\Omegadm = \Omegasdm+\Omegadcdm$. In such models, the important parameters are the fraction of decaying DM, $\fdcdm= \Omegadcdm/\Omegadm$, and its decay rate $\Gammadcdm$, which we report in units\footnote{For translation with other works making use of km s$^{-1}$Mpc$^{-1}$, we recall that 1 km s$^{-1}$Mpc$^{-1}$ = 1.02$\times 10^{-3}$ Gyr$^{-1}$.} of Gyr$^{-1}$. When $\fdcdm$ is small enough, the decay rate can in principle be very large, leading to different comsological effects than in the fully-decaying DM model.\\

\subsection{Impact of Dark Matter decay on the CMB}\label{sec:effectsCMB}
In order to show the effect of varying $\Gammadcdm$ on the CMB angular power spectra, some choice must be made about what to keep constant.
Here we choose to set all the parameters so that the early cosmological history stays the same as in the standard $\Lambda$CDM (until the decay starts).
To do so, as in ref.~\cite{Audren14}, we compare decaying models with a given value of $\omegadm^{\rm ini} = \omegasdm+\omegadcdm^{\rm ini}$ with $\Lambda$CDM models having the same $\omega_{\rm cdm}$. We also keep fixed the baryon abundance $\omega_{\rm b}\equiv \Omega_b\,h^2$, the amplitude of primordial perturbation accounting for the late-time absorption $\exp(-2\tau_{\rm reio}) A_s$, the index of the primordial perturbation spectrum $n_s$, the redshift of reionisation $z_{\rm reio}$\footnote{We follow the standard parameterisation of reionisation with a sharp hyperbolic tangent step in the ionisation fraction, centered at $z_{\rm reio}$ and of width  $\Delta z_{\rm reio}$ = 0.5.} and the angular size of the sound horizon $\theta_s$. This choice implies that if the decay happens at late times, the small-scale / high-$\ell$ part of the CMB spectra, influenced mainly by the early evolution, should be preserved up to lensing effects. We fix all parameters to their best-fit value for Planck 2015 TT,TE, EE+low-P \cite{Planck15}: \{$\theta_s$=1.04077, $\omegadm^{\rm ini}$ or $\omega_{\rm cdm}$ = 0.1198, $\omega_{\rm b}$ = 0.02225, $\ln(10^{10}A_s\exp(-2\tau_{\rm reio}))$=1.882, $n_s$=0.9645, $z_{\rm reio}$=9.9\}. Note that by fixing these parameters and varying $\Gammadcdm$ and $\fdcdm$, we expect to obtain different values for the actual $\omega_{\rm dm}$ today, for $H_0$ and for $\tau_{\rm reio}$.
Hence $\Gammadcdm$ and $\fdcdm$ can in principle be constrained by the data and lead to different predictions for $H_0$, $\sigma_8$ and $\Omegam$, that may either increase or reduce the tension with astronomical data, see our comments  in section \ref{sec:discrepancies}.\\
For illustration purposes, let us set the fraction of decaying DM to $20\%$ and compute the TT and EE CMB power spectra for three typical decay rates $\Gammadcdm=0.1,10^3,10^6$ Gyr$^{-1}$.
The spectra and their residuals are plotted in Fig.~\ref{fig:ComparaisonSpetra1}, together with the (binned) cosmic variance uncertainty.
These rates were chosen to highlight three qualitatively different regimes:
\begin{figure}
\centering
\includegraphics[scale=0.51]{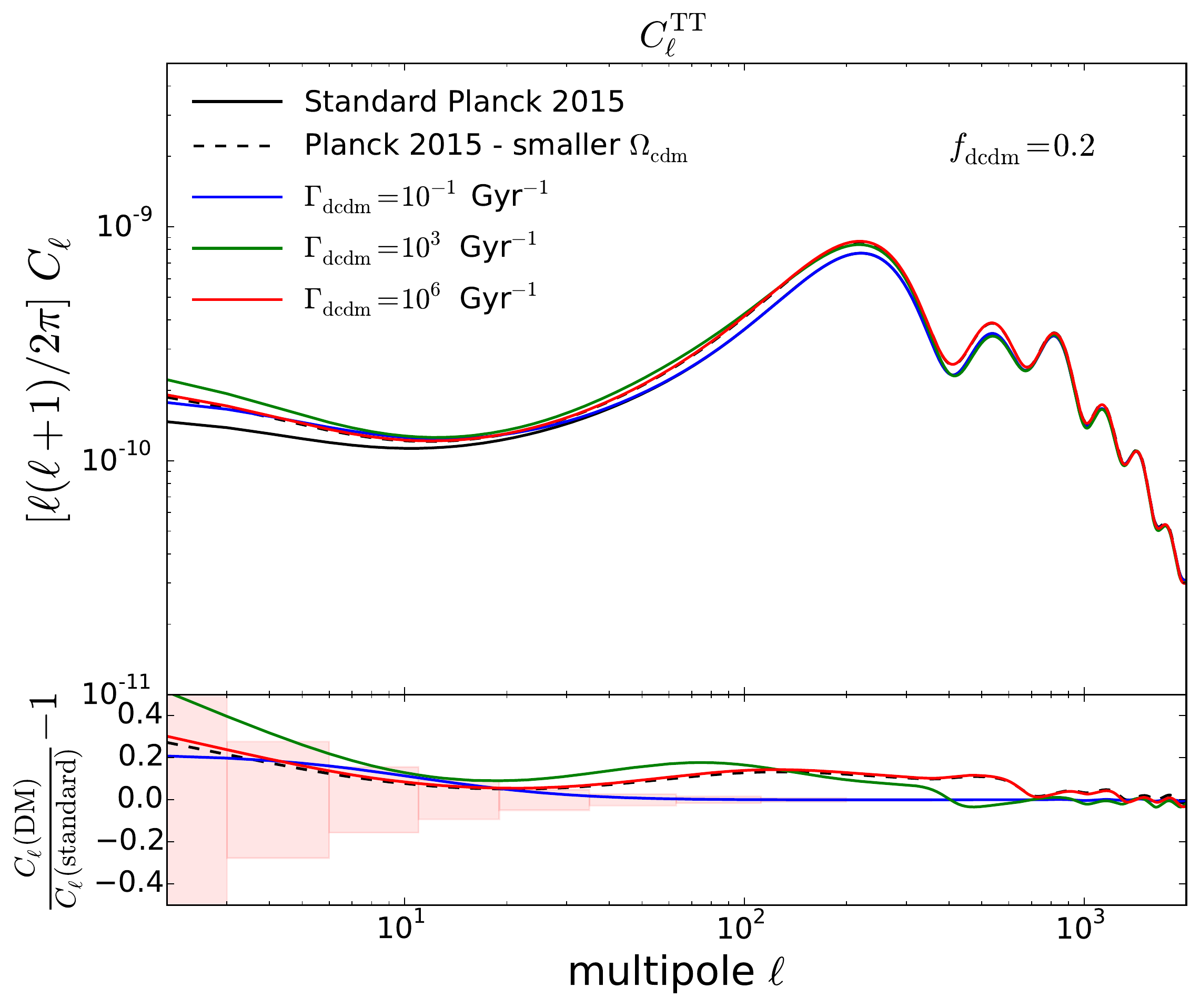}\\
\includegraphics[scale=0.51]{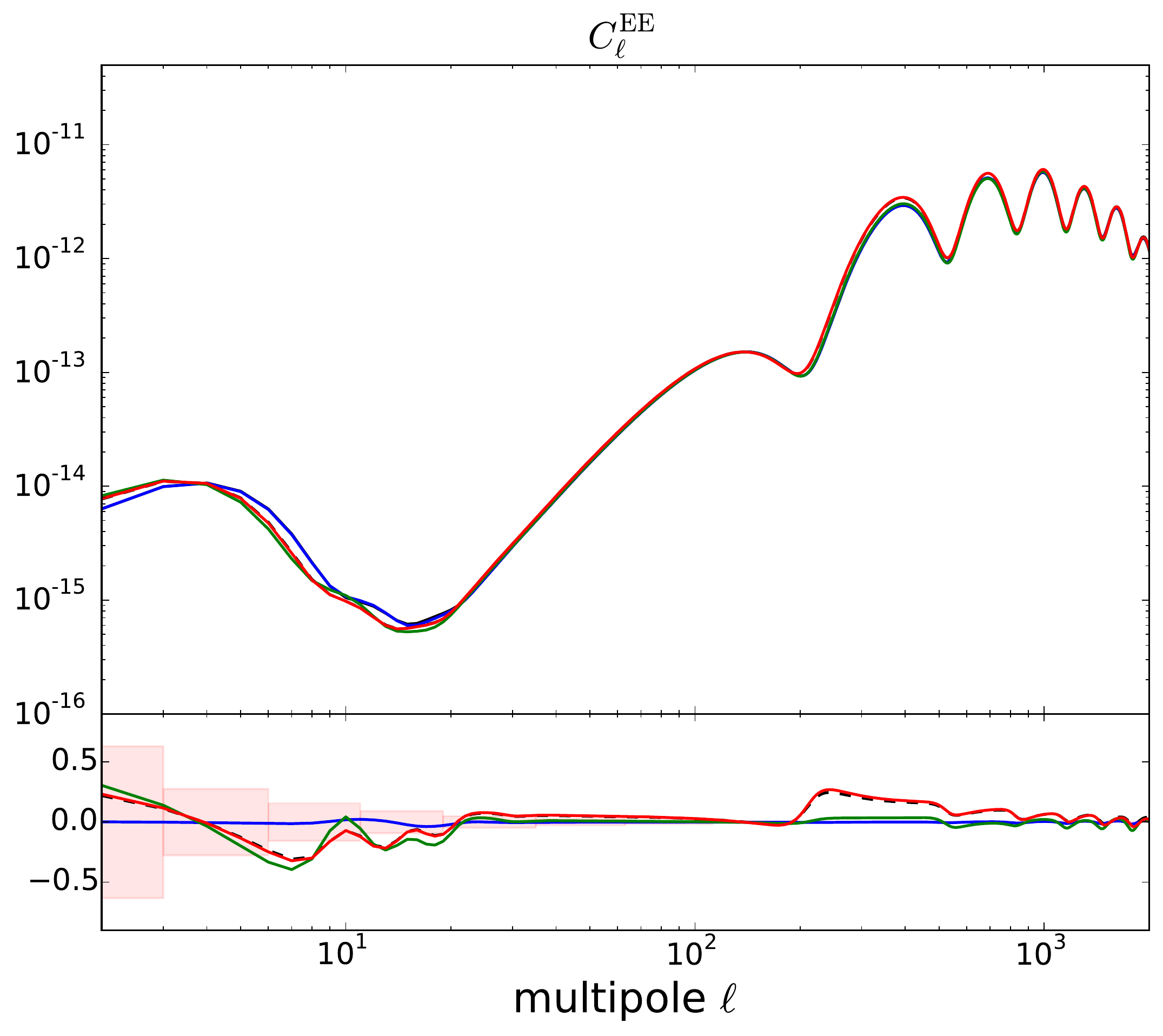}
\caption{\label{fig:ComparaisonSpetra1} Comparison of the lensed TT (top) and EE  (bottom) power spectra for several decaying DM lifetimes and a fixed abundance $f_{\rm dcdm} = 0.2$. Boxes show the (binned) cosmic variance uncertainty.}
\end{figure}

\begin{enumerate}
\item[\textbullet]The first regime (Fig. (\ref{fig:ComparaisonSpetra1}), blue curve, $\Gammadcdm = 0.1$ Gyr$^{-1}$) corresponds to a DM decaying well {\em after} recombination.  At the background level, the decrease of $a^3 \rho_{\rm m}$ with time would tend to increase the angular diameter distance to the last scattering surface. Since we are fixing $\theta_s$, this effect gets compensated by an increase in $\rho_\Lambda$ and $\Omega_\Lambda$, which in turn shifts the matter-$\Lambda$ equality redshift $z_\Lambda$ to higher values, and enhances the Late Integrated Sachs-Wolfe (LISW) effect, as can be seen in the low-$\ell$ part of the curve. On top of this, at the perturbation level, the modification of the DM density at late times generates a damping of the metric fluctuations through the Poisson equation, which further contributes to the LISW enhancement. Finally, the amount of lensing is significantly reduced. The consequences at high-$\ell$ (contrast between minima and maxima and damping tail) are hardly visible by eye on the figure, but the data is sensitive to this effect.
\\
In the EE spectrum, the different late-time evolution of $\rho_{\rm m}+\rho_\Lambda$ induces a peculiar pattern around $l\sim10$.  Indeed, the reionisation optical depth $\tau_{\rm reio}$ \footnote{defined using $\tau(z) = \int_0^z\sigma_Tx_e(z')n_H(z')\frac{dt}{dz'}dz'$, with $\frac{dt}{dz'}=-((1+z')H(z'))^{-1}$.} is an integrated quantity along $z$, and therefore also feels modification in the background expansion.
  Hence, for a fixed parameter $z_{\rm reio}$, the value of the optical depth $\tau_{\rm reio}$ and the details of the reionisation history slightly depend on $\Gammadcdm$. For a fixed product $\exp{(-2\tau_{\rm reio})} A_s$, this has consequences in the low-$\ell$ part of the EE (and also TT) spectra. However, this effect is unimportant because for a given (small) $\Gammadcdm$ the LISW effect is stronger. For the maximal allowed values of $\Gammadcdm$, reionisation effects remain below cosmic variance.\\
On the other hand, lensing impacts the C$_{\ell}^{EE}$ more strongly than C$_{\ell}^{TT}$ \cite{Galli:2014kla}. Hence, the high-$\ell$ part of the polarisation spectrum is expected to help for better constraining the lifetime and fraction of the dcdm component. This statement will be explicitely checked in section \ref{sec:results}.
\\
In summary, in this regime, the DM lifetime is probed through the LISW and lensing effect. We can further distinguish two sub-cases depending on the value of $\Gammadcdm$:\\
{\it (i)} for $\Gammadcdm \gtrsim  H_0 \sim 0.7$~Gyr$^{-1}$, most of the decaying DM has disappeared nowadays, and even before the redhsifts range $0<z<3$ which is important for the LISW and lensing effects. So in this regime we expect to get bounds on $\fdcdm$ nearly independent of $\Gammadcdm$.\\
{\it (ii)} for very small $\Gamma_{\rm dcdm}\lesssim H_0$, only a fraction of dcdm had time to disappear.
Factorizing out the expansion term, it is possible to write the evolution of the background DM density as
\begin{eqnarray}
\Omegadm &= & \Omega_{\rm sdm}+\Omega_{\rm dcdm} \nonumber\\
& = & (1-\fdcdm)\Omegadm^{\rm ini}+ \fdcdm\exp(-\Gammadcdm t)\Omegadm^{\rm ini} \nonumber\\
& = & (1-\fdcdm)\Omegadm^{\rm ini}+ \fdcdm[1-\Gammadcdm t+\mathcal{O}((\Gammadcdm t)^2)]\Omegadm^{\rm ini} \nonumber\\
& = & [1-\fdcdm\Gammadcdm t+\mathcal{O}((\Gammadcdm t)^2)]\Omegadm^{\rm ini}~.
\end{eqnarray}
In the limit $\Gamma_{\rm dcdm}\ll H_0$, terms of order two or higher can be neglected, and the remaining relevant parameter is simply $\xi_{\rm dcdm}\equiv f_{\rm dcdm}\Gamma_{\rm dcdm}$: multiplied by the age of the universe, it fully encodes the fraction of DM density which decayed into dark radiation until today. Hence this should be the quantity constrained by the data. \item[\textbullet]The second regime (Fig. (\ref{fig:ComparaisonSpetra1}), green curve, $\Gammadcdm = 10^3$ Gyr$^{-1}$) is an intermediate regime for which the unstable DM component would start to decay {\em around} the recombination epoch and has fully disappeared by now. In the CMB power spectra, one can see, on top of previously described effects, the impact of a bigger Early Integrated Sachs Wolfe (EISW) effect, since the metric terms are further damped due to the DM decay. The affected multipole $\ell$ depends on the DM lifetime whereas the amplitude of the variation depends on the fraction allowed to decay. The angular power spectra are sensitive to the two independent parameters $f_{\rm dcdm}$ and $\Gammadcdm$.
\item[\textbullet] In the third case, for very large $\Gammadcdm$ (Fig. (\ref{fig:ComparaisonSpetra1}), red curve, $\Gammadcdm = 10^6$ Gyr$^{-1}$), the unstable component of DM has decayed well {\em before} recombination, and eventually even before matter-radiation equality. One can see the admixture of previous effects together with a bigger Sachs Wolfe term, because in models with smaller $\Omegadm$, the growth of potential wells is reduced and therefore their amplitudes at the time of last scattering is smaller. Eventually, there is also a modification of the gravitationally driven oscillations that affect modes well inside the sound horizon during radiation domination, leading to small wiggles at high-$\ell$'s (visible even in the unlensed spectrum ratios). Finally, although not very pronounced in our case, if the matter radiation equality is shifted, the different expansion evolution would result in a different sound horizon at decoupling. Since we have fixed the peak scale, the code has to adapt the angular diameter distance at recoupling $d_A(z_{\rm rec})$ by adjusting  $\Omega_\Lambda$. However, $d_A(z_{\rm rec})$ also enters the diffusion damping angular scale, resulting in a small decrease in the slope at high-$\ell$'s.\\
Let us note an important point in this regime. The spectra of a model with early decaying DM are very close to those of a stable $\Lambda{\rm CDM}$ model,
with a different value of  $\Omegadm$ corresponding to the density after the decay (compare the red and black-dashed curves in figure \ref{fig:ComparaisonSpetra1}).
Hence models of this type must be allowed, provided that the final DM density is close to the best-fit value for $\Lambda{\rm CDM}$. Therefore we expect that the constraints on $\fdcdm$ start to relax as one moves to shorter and shorter lifetimes, accompanied however by an increase in the value of $\Omegadcdm^{\rm ini}$. Differents regimes are also expected, depending if the DM decays before or after matter radiation equality. Indeed a shift of $z_{\rm eq}$ induces very peculiar effects, see section \ref{sec:results}.
\end{enumerate}

\subsection{Impact of the decaying Dark Matter on the matter power spectrum}
\label{sec:MatterPowerSpectrum}
Let us now discuss the effects on the matter power spectrum, an essential step in view of including data on Large Scale Structure (LSS).
We adopt the same strategy as for the description of the CMB power spectra, i.e. we fix 
\{$\theta_s$, $\omegadm^{\rm ini}$, $\omega_{\rm b}$, $A_s\exp(-2\tau_{\rm reio})$, $n_s$, $z_{\rm reio}$\} to the Planck 2015 TT, EE, TE+low-P best fit parameters (which means that we consider a fixed early cosmological history).

\begin{figure}[!htp]
\centering
\includegraphics[scale=0.44]{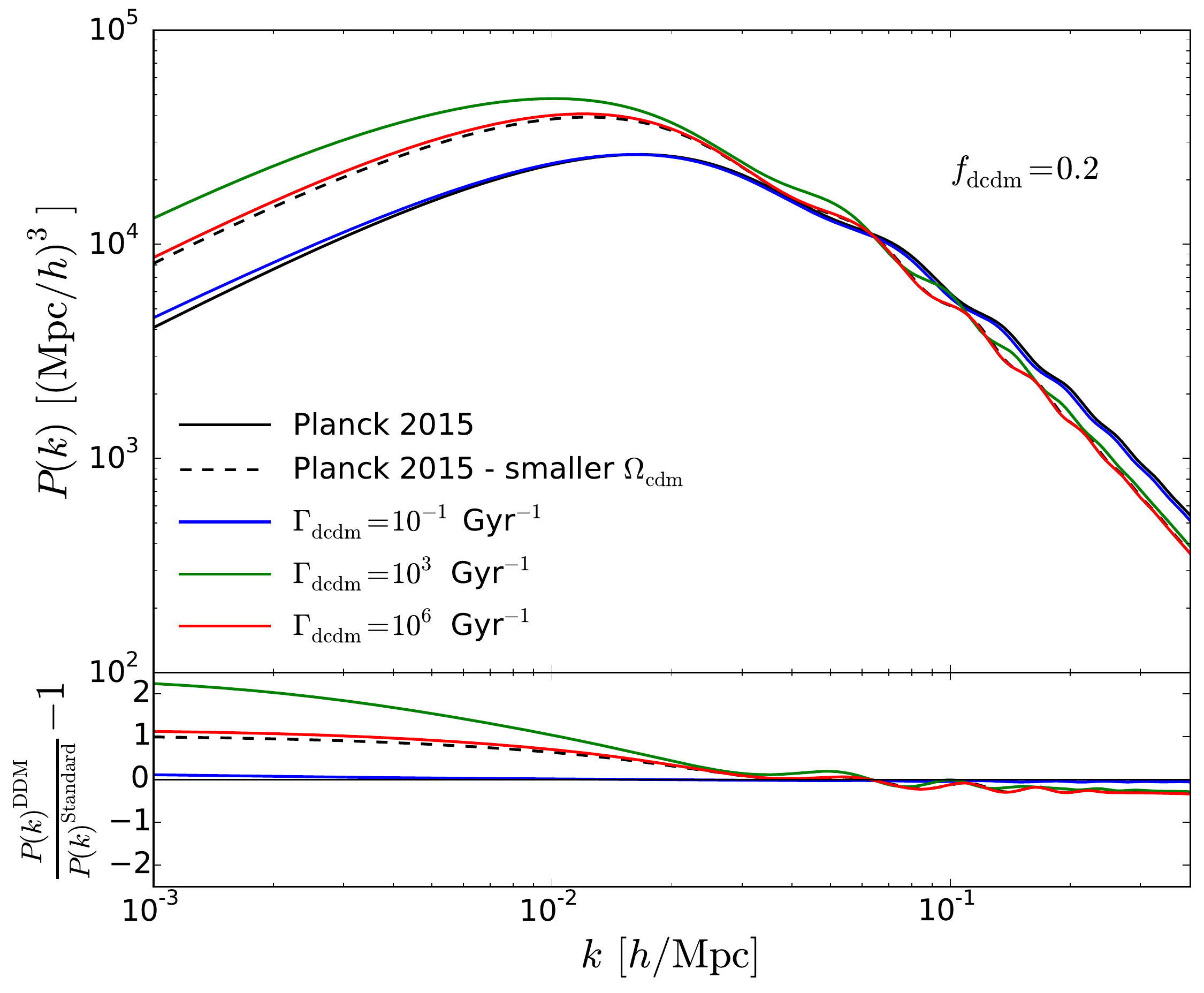}\\
\includegraphics[scale=0.44]{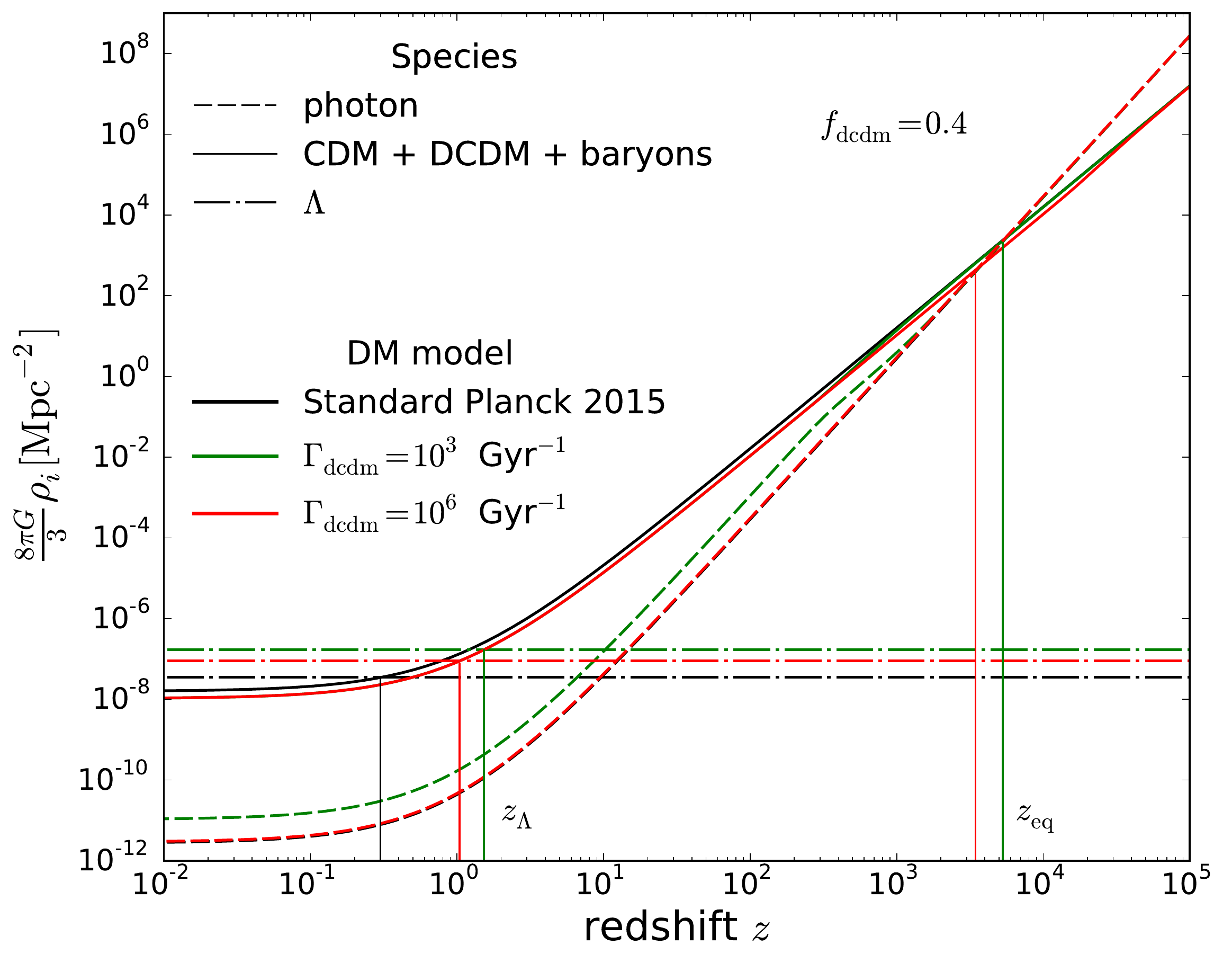}
\caption{\label{fig:ComparaisonMatterPowerSpectra} Top panel $-$ Comparison of the matter power spectrum for several decaying  DM lifetimes and a fixed abundance $f_{\rm dcdm} = 0.2$. Bottom panel $-$ Evolution of the background densities for two different models of DM with $f_{\rm dcdm} = 0.4$, compared to the standard $\Lambda$CDM model. The vertical lines indicates the value of the matter-radiation and matter-$\Lambda$ equalities in each model. Note that matter-radiation equality happens at the same redshift in the standard $\Lambda$CDM and the DCDM model with $\Gammadcdm=10^3$ Gyr$^{-1}$, therefore lines are superimposed.}
\end{figure}

As for the CMB power spectra, we show in Fig.~\ref{fig:ComparaisonMatterPowerSpectra} (top panel) the impact of DM decay for three different regimes:
\begin{enumerate}
\item[\textbullet]For lifetimes comparable or longer than the age of the Universe (blue curve), the impact on the matter power spectrum is relatively small. Mostly, one finds a small shift of the matter power spectrum towards larger scales/smaller wavenumbers. Indeed, the DM decay shortens slightly the matter dominated era\footnote{the DM decay at late time implies a smaller conformal age of the universe $\tau_0$, while the time of radiation--matter equality $\tau_\mathrm{eq}$ is fixed. This implies smaller ratios $\tau_0/\tau_\mathrm{eq}$ and $k_\mathrm{eq}/(a_0 H_0)$.}. Hence the ratio $k_\mathrm{eq}/(a_0 H_0)$ governing the location of the maximum in the matter power spectrum\footnote{We recall that $k_{\rm eq}$ is defined through the relation $k_{\rm eq}\equiv a_{\rm eq}H_{\rm eq}$ where the subscript ``eq'' stands for matter-radiation equality. The horizontal axis in Figure \ref{fig:ComparaisonMatterPowerSpectra} displays $k$ in units of [$h$/Mpc], with the implicit assumption that $a_0=1$. In fact, this axis represents the ratio $k/(a_0H_0)$, and the peak position is given by $k_{\rm eq} / (a_0 H_0)$ \cite{Lesgourgues:2006nd}.} is smaller. Apart from this small shift, the shape of the matter power spectrum is unchanged, because all the cosmological evolution until the time of recombination and of baryon drag is fixed. 
\item[\textbullet] For a decay starting around the time of baryon drag (green curve in Fig.~\ref{fig:ComparaisonMatterPowerSpectra}), the impact is much stronger and can be decomposed in three effects:\\
(i) we observe the same shift to larger scales/smaller wavenumbers as in the previous case, occurring for the same reason, but much more pronounced. \\
(ii)  the amplitude on large scales is bigger. Indeed, the amplitude of the matter power spectrum $P(k)$ (expressed in units of $h^{-3}$Mpc$^3$ versus $k$ in $h$Mpc$^{-1}$) on scales $k\ll k_{\rm eq}$
depends on the primordial spectrum multiplied by $(g(a_0,\Omegam)/\Omegam)^2$ \cite{Lesgourgues:1519137}. Here $g(a,\Omega)$ is the function expressing how much the growth rate of structures $D(a)$ is suppressed during $\Lambda$ domination: it is the ratio $D(a)/a$ normalised to one before $\Lambda$ domination, and in a flat FLRW universe, it depends only on $a$ and $\Omega_m=1-\Omega_\Lambda$. So $g(a_0,\Omegam)$ is a simple function of $\Omegam$, growing with $\Omegam$ and reaching one for $\Omegam=1$. However 
$g(a_0,\Omegam)$ does not increase as much as $\Omegam^2$, so the ratio  $(g(a_0,\Omegam)/\Omegam)^2$ decreases with $\Omegam$.
In the dcdm model that we are considering now, the decay leads to a smaller matter density at late times, and to maintain a constant angular diameter distance to recombination and a constant $\theta_s$, one needs to increase $\Lambda$. Hence $\Omegam$ is smaller and the large--scale power spectrum is enhanced. \\
(iii) the small-scale power spectrum is suppressed and has a different shape. Indeed we are considering a fixed value of $\Omega_{\rm b}$, while $\Omega_{\rm dm}$ is smaller in the dcdm model. This means that the ratio $\Omega_{\rm b}/\Omega_{\rm dm}$ is bigger. Since baryons are coupled to photons until the baryon drag epoch, a larger 
$\Omega_{\rm b}/\Omega_{\rm dm}$ implies a strong small-scale suppression and larger Baryon Acoustic Oscillations (BAOs). There is also a slight shift of the BAO phase, due to the change in the value of the sound horizon at baryon drag. 
\item[\textbullet] The effects described in the previous regime do not keep increasing monotonically when the lifetime decreases. On the contrary, for a decay happening well before baryon drag, the impact of the decay becomes smaller (red curve in Fig.~\ref{fig:ComparaisonMatterPowerSpectra}). This is related to the variation of $\Omegam$ with the dcdm lifetime, when $\theta_s$ and all other cosmological parameters are kept fixed. As long as the DM lifetime is longer than the recombination time, it impacts $\theta_s=d_s/d_A$ only through the angular diameter distance to the last scattering surface $d_A$, because the expansion history is different after decoupling. Once the DM lifetime becomes smaller than the recombination time, the time of radiation--matter equality changes, and $\theta_s$ is also impacted through the value of the sound horizon at decoupling $d_s$. In all cases $\Omega_\Lambda$ is automatically adjusted in order to get the same $\theta_s$, but in a non-monotonic way. This is true also for $\Omegam=1-\Omega_\Lambda$, which
first decreases with $\Gammadcdm$ and then increases. This explains why in the left panel of 
Fig.~\ref{fig:ComparaisonMatterPowerSpectra}, the amplitude of the small-scale matter power spectrum first goes up with $\Gammadcdm$, and then goes down. 
The bottom panel helps to understand what is going on with the background evolution in the different models. On top of this effect, the different value of the sound horizon at decoupling leads to further shift in the BAO phase.

Finally and as expected, in the limit of a very small DM lifetime, the matter power spectrum asymptotes to the limit of a stable CDM model with a smaller $\Omegadm$ (dashed black curve), corresponding to the DM density of the dcdm model after full decay.

\end{enumerate}

\subsection{Potential degeneracy with the neutrino mass}

In the past literature, the DM lifetime has been found to be partially degenerate with several other parameters.
For instance, ref.~\cite{Audren14} studied extensively the degeneracy between $\Gammadcdm$, the curvature of the universe and the tensor mode amplitude. They found some degeneracy at the level of primary CMB anisotropies, fortunately broken by CMB lensing and matter power spectrum data.
Other authors, for instance ref.~\cite{Berezhiani:2015yta, Enqvist:2015ara}, have found some correlation between $\Gammadcdm$ and ($H_0$, $\sigma_8$), as expected from the previous discussion. They pointed out that this could be helpful in resolving tensions with low redshift astronomical data. In the next section, we scrutinize this claim.

Besides, in principle, one could expect a degeneracy between the DM decay and neutrino mass effects. This has been paid virtually no attention till now, hence we discuss it in the following.
Neutrinos are relativistic at early time, and then experience a non-relativistic transition. Schematically, one could say that decaying DM goes the opposite, since the model features a non-relativistic species which, through its decay into dark radiation, undergoes a sort of ``relativistic transition''. For a fixed value of neutrino mass, one can play with the fraction of dark matter that decays in order to cancel exactly the increase in $\Omegam$ coming from the neutrino sector. At the same time, one can adjust the time of the decay to match the neutrino non-relativistic transition time.

We wish to check this simple statement by looking at the power spectra.
As it is well known, the effects of the neutrino mass on the CMB power spectra highly depends on the epoch at which the non-relativistic transition occurs, and therefore on the value of the neutrino mass (see e.g. Ref.~\cite{Lesgourgues:1519137,Lesgourgues:2006nd}
 reviews).
The larger it is, the earlier the transition happens, eventually even before matter radiation equality for $m_\nu > 1.5$ eV. Such high masses are completely ruled out by observations\footnote{We anticipate that there is no need to put this conclusion to renewed scrutiny, since the potential degeneracy between DM decay and neutrino mass effects is in fact lifted given current data.} and our discussion will be restricted to masses for which the non-relativistic transition happens after recombination, $m_\nu<0.6$~eV.

For such masses, neutrinos simply affect the CMB through post-recombination effects: EISW and LISW, lensing, and a modification of the angular diameter distance to the last scattering surface. We want to check whether we can cancel these effects with some appropriate amount of DM decay. Hence we compare different 
models with common parameters
\{$\theta_s$, $\omegadm^{\rm ini}$ or $\omegadm$, $\omega_{\rm b}$, $A_s$, $n_s$, $\tau_{\rm reio}$\}, three degenerate massive neutrino species,
and different values of $M_\nu=3 m_\nu$, $\Gammadcdm$ and $\fdcdm$. Note that all these models share the same cosmological evolution until the time at which either DM decays, or neutrinos become non-relativistic (in particular, for $m \ll $ 1.5~eV and a lifetime much bigger than the time of radiation--matter equality, $\tau_\mathrm{eq}$ is the constant). Moreover, by fixing $\theta_s$, we remove one of the potential effects of neutrino masses listed above. Fixed values of $\theta_s$ are obtained by adjusting $\Omega_\Lambda$ in each model, which only leaves a signature at the level of the late ISW effect.

\begin{figure}[!ht]
\centering
\includegraphics[scale=0.34]{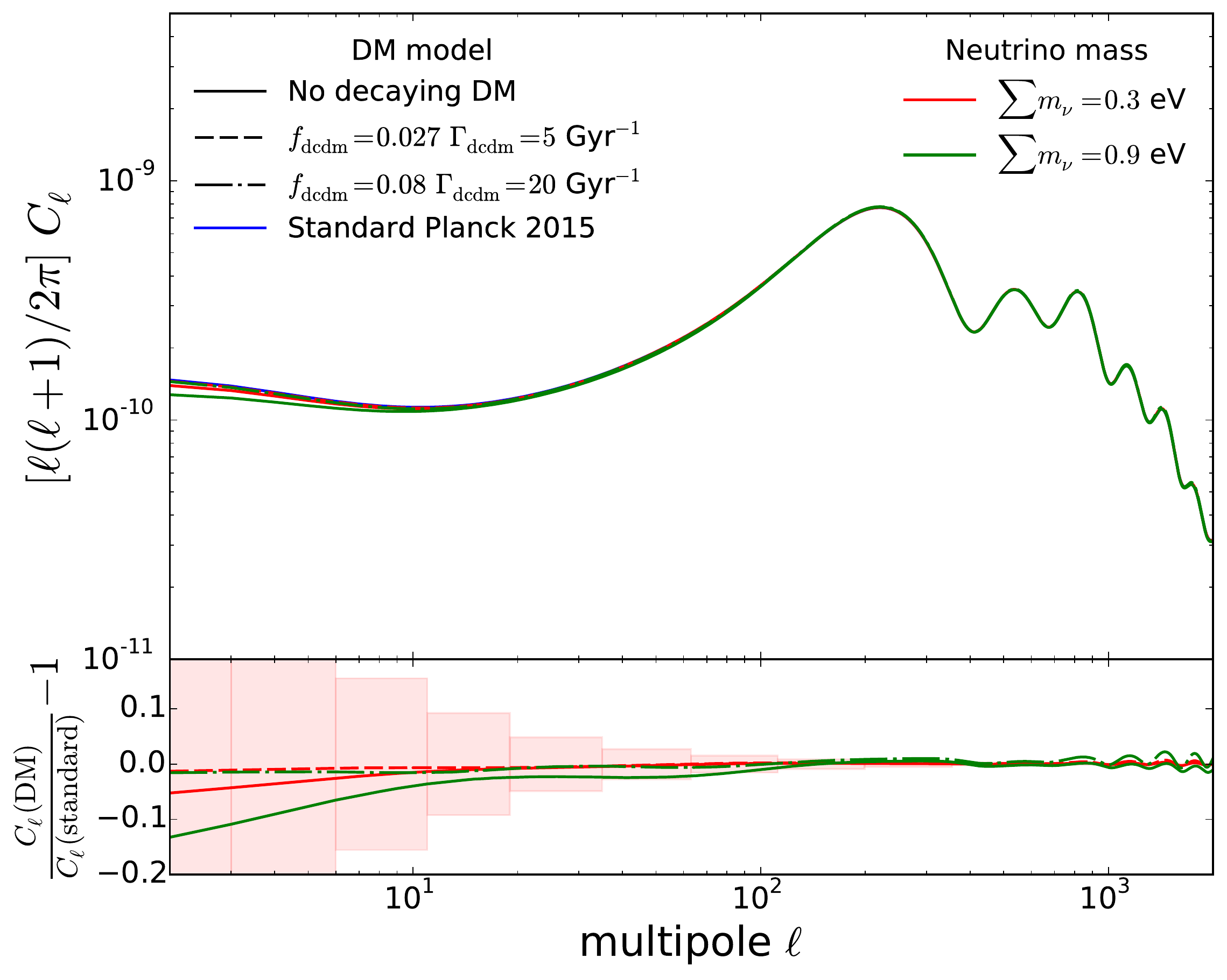}\\
\includegraphics[scale=0.34]{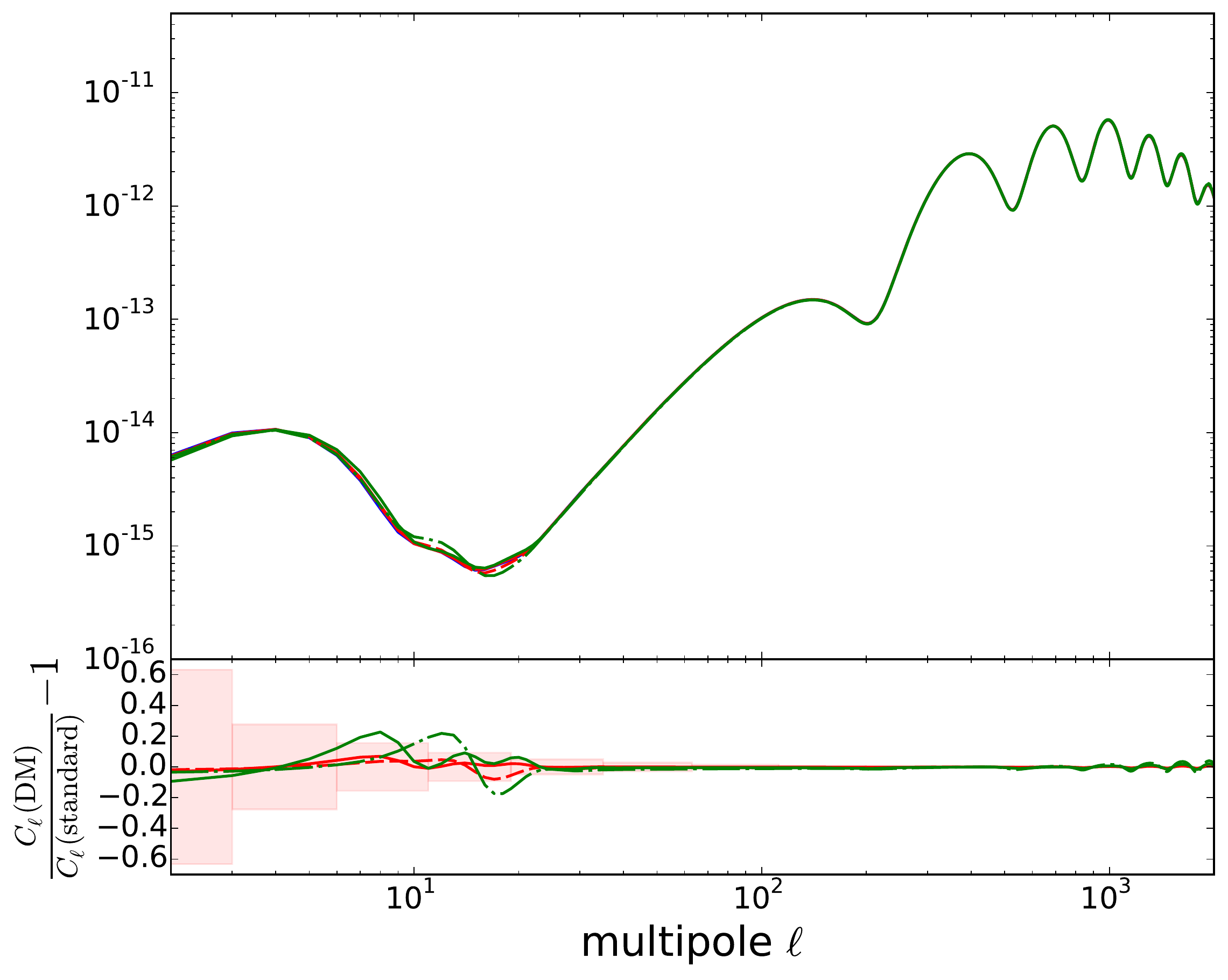}\\
\includegraphics[scale=0.34]{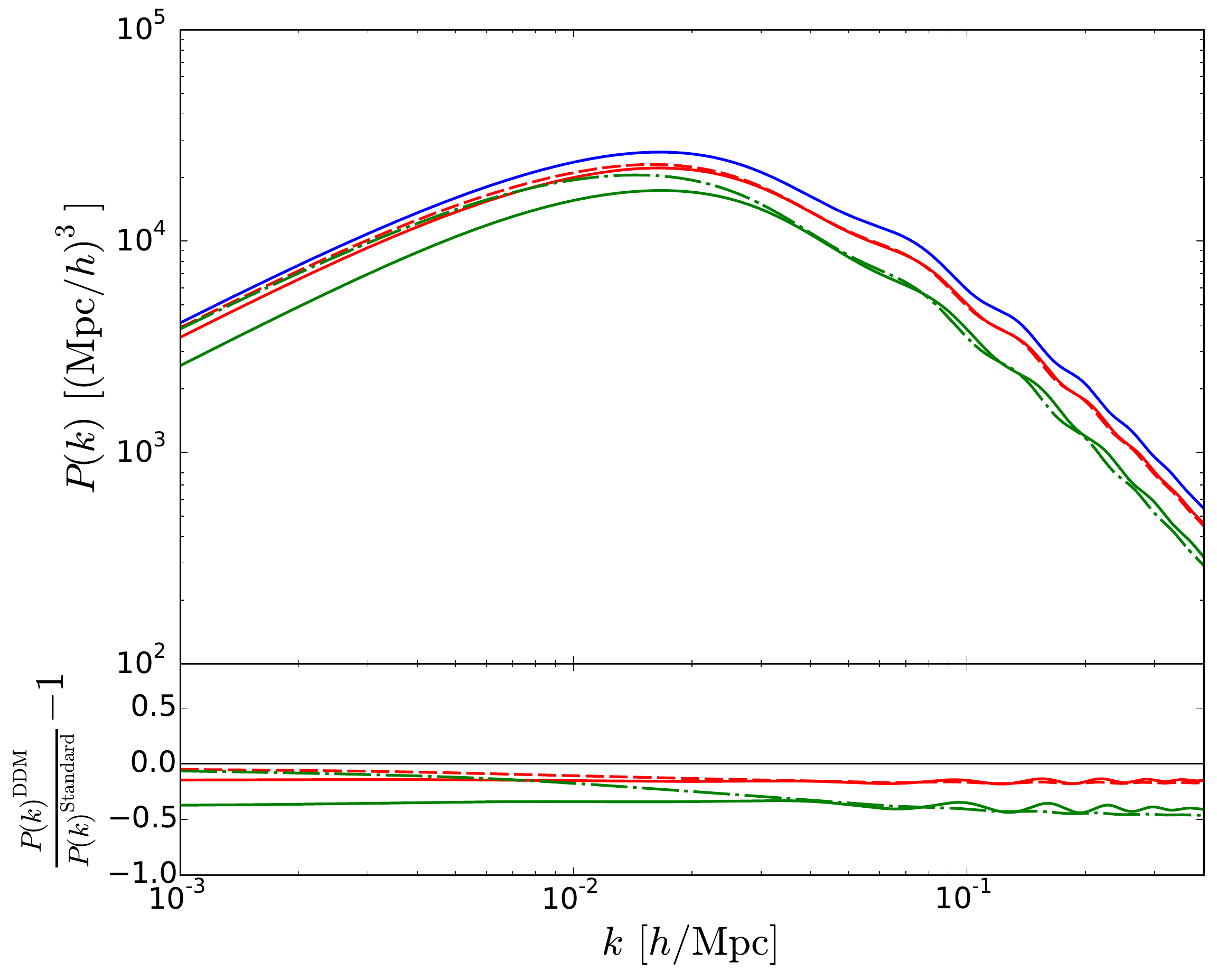}
\caption{\label{fig:DegeneracyNeutrinoMass} Comparison of the lensed TT (top), EE (middle) and matter (bottom) power spectra for several decaying DM lifetimes and several neutrino masses. The value $\fdcdm=0.027$ (resp. 0.08) has been chosen in order to 
compensate the effect of $M_\nu=0.3$~eV (resp. 0.9~eV) and obtain the same total dark matter density today, $\omegam$. The inverse of $\Gammadcdm$ has been adjusted to the time of the neutrino non-relativistic transition. Boxes show the (binned) cosmic variance uncertainty.}
\end{figure}

\begin{itemize}
\item {\it CMB temperature spectrum}.
In Fig.~\ref{fig:DegeneracyNeutrinoMass}, top panel,  one can check that while varying the neutrino mass with only stable CDM, the lensed TT spectrum is only affected in the late ISW region (depletion for $\ell\leq20$), early ISW region (depletion for $20 \leq \ell \leq200$) and lensing region (small wiggles for $\ell \geq 1000$). Then, for a given neutrino mass, we increased $\fdcdm$ in order to obtain roughly the same total $\omegam=\omega_{\rm b}+\omegasdm+\omegadcdm+\omega_\nu$ today, and choose a value of $\Gammadcdm$ cancelling as much as possible the neutrino effects. This allows to counter--act neutrino masses at the level of the unlensed spectrum: the ISW effects nearly disappear. However, the lensing effects remain at large $\ell$.

\item {\it CMB polarisation spectrum}.
In the EE spectrum (Fig.~\ref{fig:DegeneracyNeutrinoMass}, middle panel), the degeneracy is not as effective as one could expect. This is mainly because the EE spectrum is very sensitive to modifications in the reionisation history at late time. As already argued in section \ref{sec:effects}, even for a fixed thermal history and ionisation function versus redhsift $x_e(z)$, changes in the background densities affect the redshift--to--time relation $z(t)$. The effect of neutrinos becoming non--relativistic and of decaying DM can never exactly cancel each other in the expansion history $a(t)$, hence $x_e(t)$ is always slightly different, and this can be seen in  the EE spectrum, for $\ell \leq 30$ (i.e., for modes entering the Hubble radius around the reionisation time). Features are present on those scales even if we try to fix $\tau_\mathrm{reio}$ instead of $z_\mathrm{reio}$. So, despite the magnitude of cosmic variance on those scales and our ignorance on the cosmic reionisation history, we expect the EE spectrum to contribute somewhat to the breaking of the degeneracy. Besides, like for temperature, the degeneracy is broken at large $\ell$ by lensing effects.

\item {\it Matter power spectrum}.
Since CMB lensing effects depend on the matter power spectrum, we now look at $P(k)$ for the same models
(bottom panel of fig.~\ref{fig:DegeneracyNeutrinoMass}). 
It is well-known that the direct effect of neutrino masses is a step--like suppression of the small-scale matter power spectrum, coming mainly from a reduction of the growth rate of CDM fluctuations in presence of a free-streaming component. This effect in the matter power spectrum is best seen  by fixing both $\Omegam$ and $\omegam$, in order to get the same behaviour of fluctuations on scales bigger than the neutrino free-streaming scale, in the regime where neutrino and cdm perturbations are equivalent. With fixed ($\Omegam$, $\omegam$), or equivalently, fixed ($\Omegam$, $h$),  one clearly sees that neutrino masses suppress $P(k)$ only on small scales.
However, in the matter power spectrum comparison presented in this section, we wish to keep the same choice as in our previous CMB spectrum comparison: namely, we fix $\theta_s$, since this quantity is very accurately measured by the CMB,
and we also fix $\omegadm$ and $\omega_{\rm b}$, in order to keep the same early cosmological evolution. In that case,
models with different neutrino masses will also have different values of $\Omegam$ and of $h=\sqrt{(\omegadm+\omega_{\rm b}+\omega_\nu)/\Omegam}$, in order to achieve the same $\theta_s$. 
Since the value of $\Omegam$ affects the amplitude of the matter power spectrum on small scales $k \ll k_\mathrm{eq}$, we do not expect to see the usual step-like suppression on small scales. 

To explain the variation of the matter power spectrum with respect to the neutrino mass when $\theta_s$, $\omega_{\rm b}$, $\omegadm$ and the primordial spectrum are fixed, we need to remember that in the $\Lambda$CDM model, the regime $k \ll k_\mathrm{eq}$ of $P(k)$ depends only on $\Omegam$, while that of the small-scale power spectrum depends only on $z_{\rm eq}$ and on the baryon fraction (this can be checked e.g. in \cite{Lesgourgues:1519137}, from equation (6.39); to obtain the dependence on $z_{\rm eq}$, on needs to eliminate $\tilde{k}_{\rm eq}$ in favour of $z_{\rm eq}$ using equations (6.32, 6.35, 6.36)). Suppose that we first increase $\Omegam$ while keeping $\omega_{\rm b}$, $\omegadm$ and $\omega_\nu$ fixed. This will suppress the power spectrum at large scales, while keeping the same amplitude on small scales, since $z_{\rm eq}$ is not changing. Now let us increase neutrino masses. We add another step-like suppression, this times acting on small scales. After these two transformations, it is not obvious if the power spectrum gets more suppressed on small or large scales.

It turns out that when the neutrino mass is varied for fixed $\theta_s$, $\omega_{\rm b}$, $\omegadm$, the amplitude of the suppression on small and large scales is nearly the same. In that case, the neutrino mass does not manifest itself as a step--like suppression, but  as a decrease in the global amplitude. In other words,  in this basis, neutrino masses are responsible for a shift in the amplitude of the matter power spectrum, while the amplitude of the CMB spectra remains constant.
This effect comes together with a shift in BAO phases, because the angular diameter distance to small redshifts is changing (unlike the angular diameter distance to recombination). 

We can now turn on DM decay with the same motivation as before: by adjusting the dcdm fraction, we can get the same total matter density $\omegam$ today. The quantity of DM decaying into radiation gets exactly compensated by the amount of neutrinos becoming non-relativistic. In that case, the background history gets much closer to that of the initial model, and we have seen that ISW effects in the CMB are also compensated. For the matter power spectrum, the story is different. We now have the same $\omegam$ and nearly the same ($H_0$, $\Omega$) in the original $\Lambda$CDM model and in the mixed DM decay + massive neutrino model. Hence the amplitude of the matter power spectrum is preserved on large scales, the BAO scale readjusted, and the usual step-like suppression caused by neutrino masses appears clearly, not masked by other large--scale effects. This is why the difference between these two models in Fig.~\ref{fig:DegeneracyNeutrinoMass}, lower panel, looks like in a canonical comparison between models with massive or massless neutrinos for fixed ($\Omegam$, $\omegam$). 
\end{itemize}

In summary, the degeneracy which is superficially present in CMB angular spectra is clearly broken at the level of the matter power spectrum, and hence also at the level of the lensed CMB spectra, even neglecting the small distinctive  features in the polarisation spectra on large angular scales.
To cross--check this conclusion, we performed some fits to the data with free $M_\nu$, $\fdcdm$ and $\Gammadcdm$ simultaneously. We found no significant correlations between these parameters, and got neutrino mass bounds extremely close to those obtained with stable DM. We can conclude that the bounds of the next sections, obtained with massless neutrinos, are very robust against the addition of neutrino masses.

\section{Application of the decaying dark matter model}\label{sec:results}
\subsection{Constraints from the CMB power spectra only}
\begin{figure}[!htb]
\centering
\includegraphics[scale=0.8]{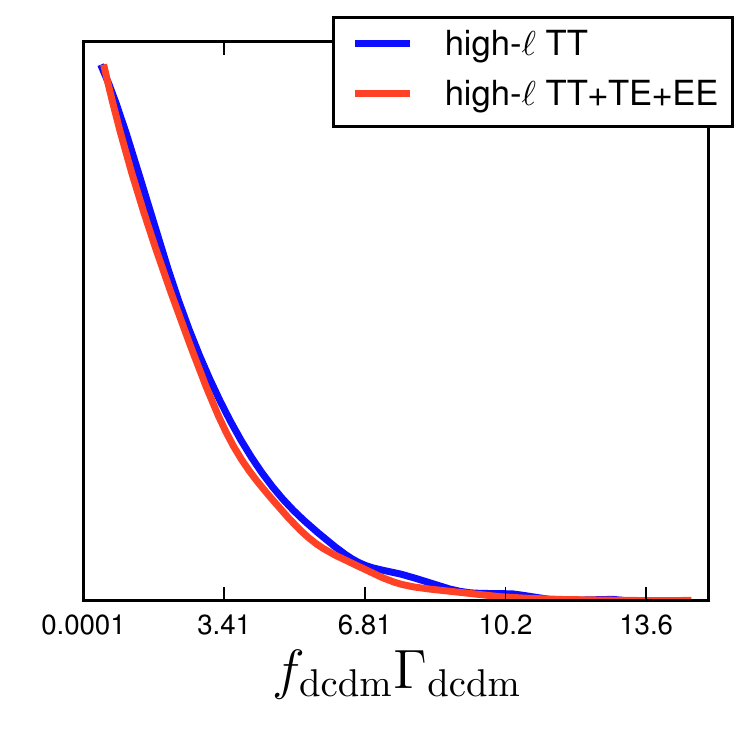}\\
\includegraphics[scale=0.65]{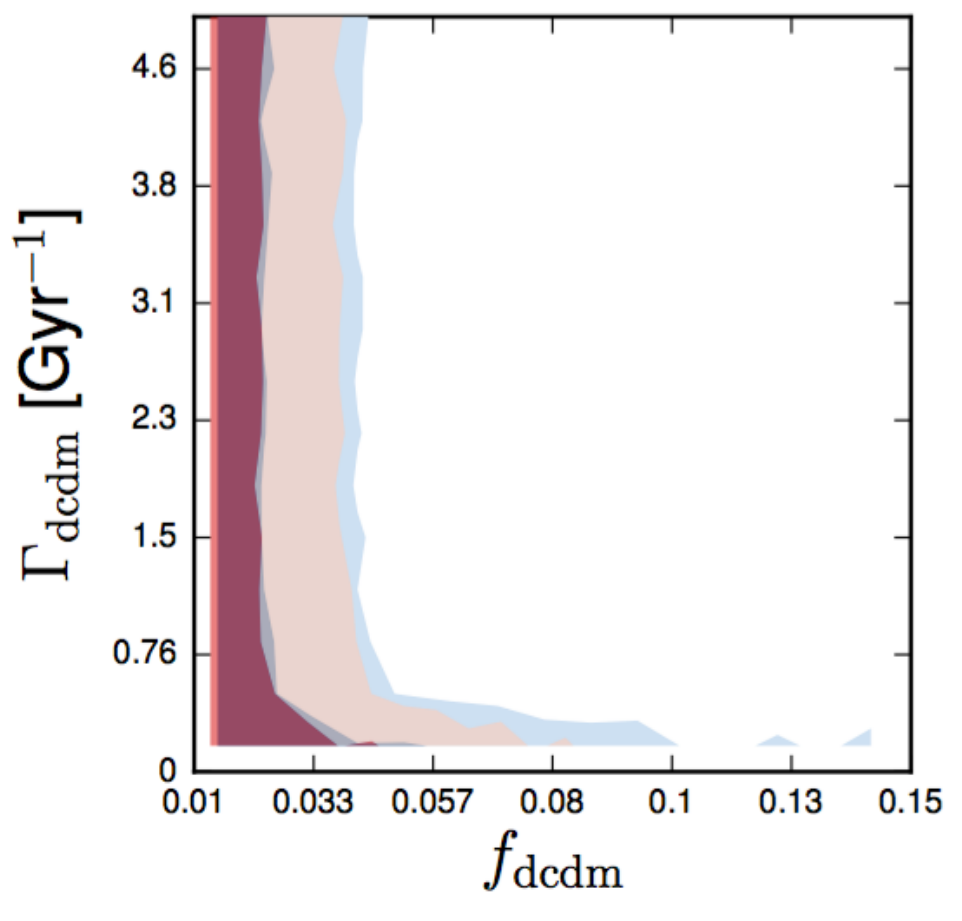}
\includegraphics[scale=0.5]{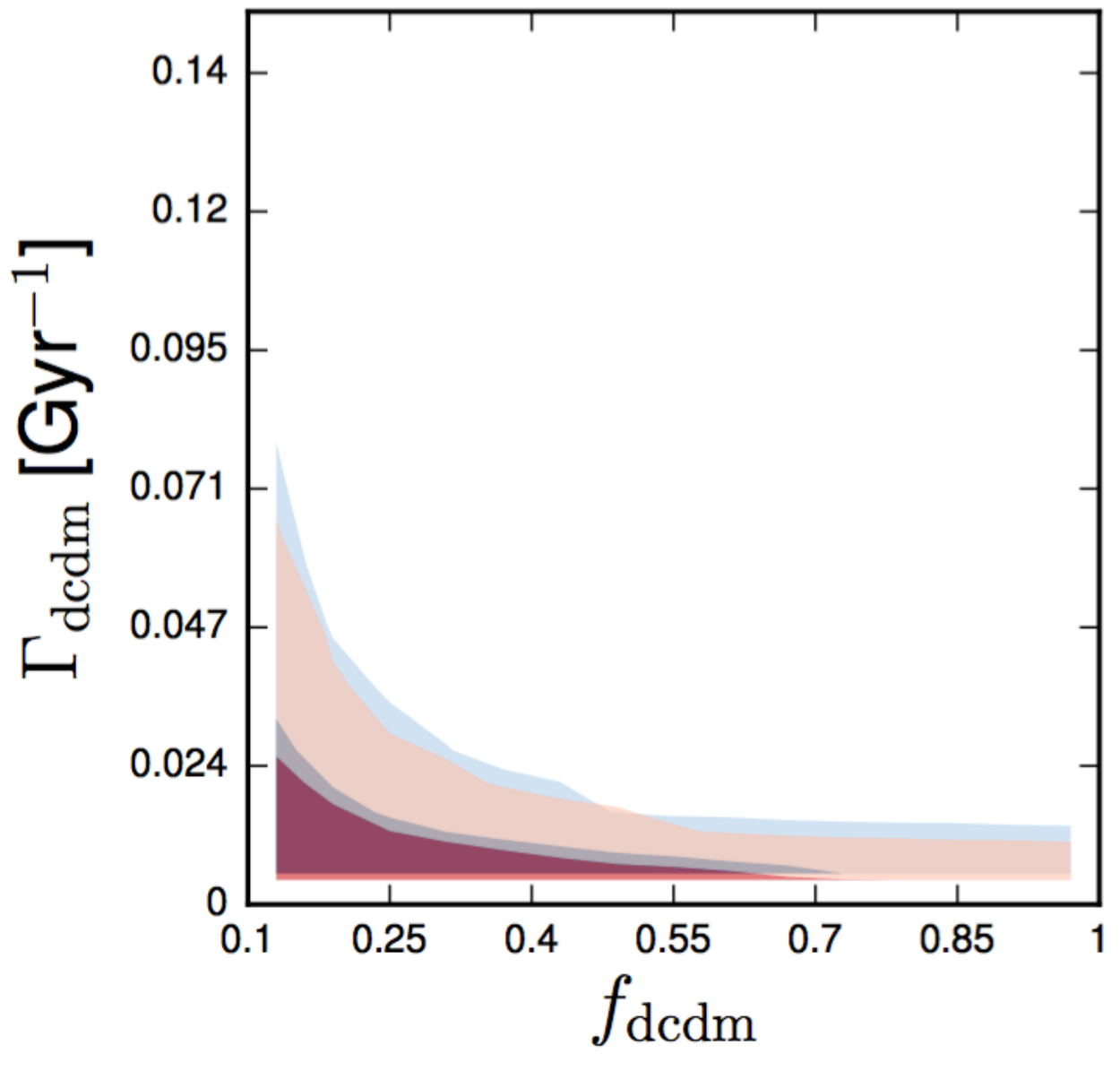}\\
\caption{\label{fig:ConstraintsCMBOnly} Constraints on the decaying dark matter fraction $\fdcdm$ as a function of the lifetime $\Gammadcdm$  in the long-lived and intermediate regime. All datasets also include CMB low-$\ell$ data from each spectrum and the lensing reconstruction. Blue (red) lines and contours refer to the case without (with) high-$\ell$ polarization data. Inner and outer coloured regions denote $1\,\sigma$ and $2~\sigma$ contours, respectively.} 
\end{figure}

\begin{figure}[!htb]
\centering
\includegraphics[scale=0.65]{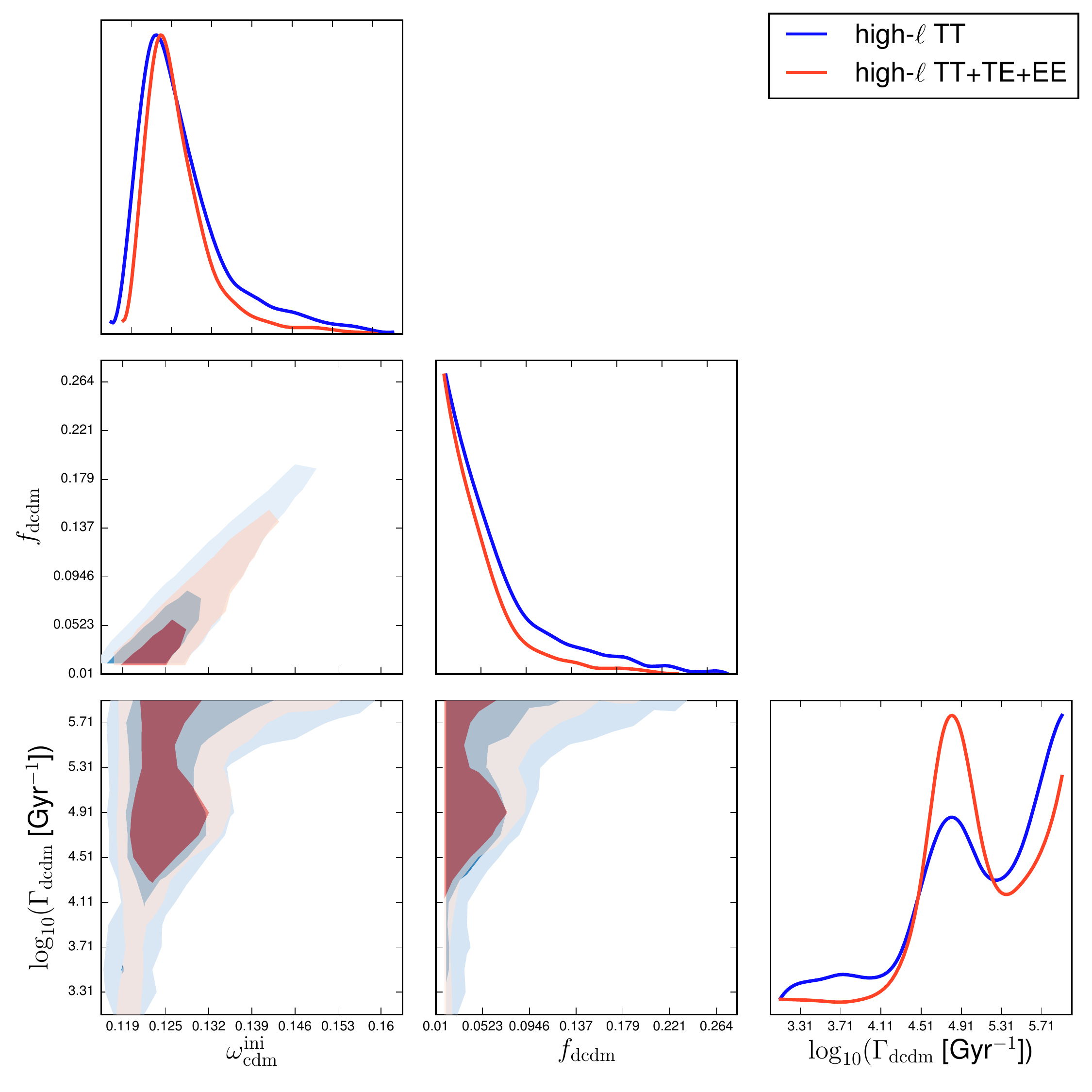}
  \caption{\label{fig:ConstraintsCMBOnly2} Constraints as in Fig.~\ref{fig:ConstraintsCMBOnly}, but for the short-lived dcdm regime. We also show how the distribution of the initial cdm density evolves when the decay rate and dcdm fraction increase.} 
\end{figure}

It is well known that $\Lambda$CDM  provides a good fit to Planck results, suggesting that bounds on decaying DM (rather than evidence in favour of it) should be  achievable by an appropriate analysis of the data. We compare the constraining power of the TT, TE and EE spectra, since the Planck collaboration made available several likelihoods, corresponding to differents data sets \cite{Planck15LKL}. 
For $\ell$'s >30, we can use data from the TT spectrum only, or use data coming from TT, TE and EE spectra at the same time.
For the small $\ell$'s however, it would not make sense to consider such decomposition since the TT spectrum alone is very weakly sensitive to $\tau_{\rm reio}$: more precisely, its effect is highly degenerate with $A_s$. We thus simply consider the combination of all data sets \{TT, TE, EE\} which allows one to break the $A_s$/$\tau_{\rm reio}$ degeneracy. Further cosmological information is encoded in the likelihood of the {\it lensing} reconstruction, which we also use in the following.
We refer to the combination of high-$\ell$'s TT + low-$\ell$'s + lensing reconstruction as Planck$^{\rm TT}$, 
and to high-$\ell$'s TT,TE,EE + low-$\ell$'s + lensing reconstruction as Planck$^{\rm TTTEEE}$. Ref.~\cite{Galli:2014kla} has found that despite their rather poor signal-to-noise ratio, the constraining power of the EE and TE spectra alone is as good as or even better than the TT alone. In fact, they found that C$_{\ell}^{TE}$ improves the determination of $\omegadm$ by 15\%. One therefore expects similar improvements when using Planck$^{\rm TTTEEE}$ with respect to Planck$^{\rm TT}$.\\
We run Monte Carlo Markov chains using the public code {\sc Monte Python} \cite{Audren12}. 
We perform the analysis with a Metropolis Hasting algorithm and assumed flat priors on the following parameters:
$$
\{\omega_b,\theta_s,A_s,n_s,\tau_{\rm reio},\omega^{\rm ini}_{\rm cdm},f_{\rm dcdm},\Gamma_{\rm dcdm}\}\,.
$$
Although not specified here for brevity, there are many nuisance parameters that we analyse together with the cosmological ones. To this end, we make use of a Choleski decomposition which helps in handling the large number of nuisance parameters \cite{Lewis:2013hha}. We consider chains to be converged using the Gelman-Rubin \cite{Gelman:1992zz} criterium $R -1<0.01$, except if specified otherwise. \\
As extensively discussed in section \ref{sec:effects}, the effects of the decay are very different depending on the lifetime of the decaying DM. Therefore, the parameter space should have a non-trivial shape: we therefore split our analysis in three different parts, corresponding to different decay epochs.
In Fig.~\ref{fig:ConstraintsCMBOnly} and Fig.~\ref{fig:ConstraintsCMBOnly2}, with blue curves and contours, we show the constraints in the $\{\fdcdm,\Gammadcdm\}$ plane for each regime from the Planck$^{\rm TT}$ dataset only.  

\begin{enumerate}
\item[\textbullet] The constraints in the long-lived dark matter regime ($\Gammadcdm \lesssim H_0$) is represented on the upper and bottom-left panels of Fig.~\ref{fig:ConstraintsCMBOnly}. In that regime, a sizable fraction of the  decaying DM component is still present today.
As anticipated in section \ref{sec:effectsCMB}, first bullet, the key parameter in that case is the product $f_{\rm dcdm}\Gamma_{\rm dcdm}$, with an exact two parameters description leading to less than 10\% differences in the bounds.  The constraints on this parameter is $f_{\rm dcdm}\Gamma_{\rm dcdm}<6.7\times10^{-3}$ Gyr$^{-1}$ at 95\% CL with Planck$^{\rm TT}$ dataset.
\item[\textbullet] We argued in section \ref{sec:effectsCMB}, second bullet, that there is an intermediate regime given roughly by $\Gammadcdm\in [10^{-1} , 10^{3}] ~{\rm Gyr}^{-1}$, for which the DM decay starts after recombination and decaying DM has totally disappeared by now.
Results for this case are shown in the bottom-right panel of Fig.~\ref{fig:ConstraintsCMBOnly}. In that case, the CMB is mostly insensitive to the time of the decay. Our runs show that in this regime, the CMB can tolerate up to 4.2\% of dcdm at 95\% CL. This is an important number, standing for the fraction of dark matter that can be converted entirely into a dark radiation after recombination, without causing tensions with the data. Although we do not show the full parameter space up to $\Gammadcdm\simeq 10^3$ ${\rm Gyr}^{-1}$, we have checked that the behaviour stays the same (this can also be inferred from the smallest values of $\Gammadcdm$ plotted in Fig.~\ref{fig:ConstraintsCMBOnly2}.)
\item[\textbullet] Finally, we show in Fig.~\ref{fig:ConstraintsCMBOnly2} the constraints applicable to the short-lived regime, $\Gammadcdm > 10^3~{\rm Gyr}^{-1}$, for which the decay happens before recombination. To accelerate the exploration of the parameter space, we scan over $\log_{10}(\Gammadcdm)$ with a flat prior. We however cut at $10^6~{\rm Gyr}^{-1}$ for obvious convergence issues, and consider chains as converged when $R-1 < 0.1$. Changing the upper bound would not change at all our conclusions. Note that with such a bound, we are also covering the region of parameter space for which the decay happens before the onset of matter domination. 
One can see a very interesting behaviour in that regime: the bound on $\fdcdm$ starts to relax, accompanied by an increase in the initial total dark matter density. In principle, cosmologies
with a large initial cold DM abundance are acceptable,  provided that the decaying DM, in excess with respect to Planck $\Lambda$CDM best fit value for $\omegadm$,
 had time to decay before recombination. In practice, we see two different regimes, responsible for non-monotonic features in the contours of Fig.~\ref{fig:ConstraintsCMBOnly2} in the $\{\fdcdm,\log_{10}(\Gammadcdm)\}$ plane. The first regime, for which the constraints relax more slowly, corresponds to decay happening mostly in between matter-radiation equality and recombination. The second regime corresponds to decay happening mostly before matter-radiation equality. The difference in the slope of the relaxation of the constraints is therefore mostly due to the fact that, in the first regime, the matter-radiation equality redshift $z_{\rm eq}$ is shifted towards earlier time. This in turn modifies the background evolution---therefore the sound horizon and the diffusion damping scale---but also directly affects the growth of both metric and density perturbations. Indeed during matter domination, the growth of matter perturbations deep inside the sound horizon is linear, whereas it is only logarithmic in the radiation domination era. Hence, a longer matter domination results in a bigger amplitude of the matter density perturbations. On the other hand, as already explained, if matter-radiation equality happens earlier, metric perturbations have more time to stabilize resulting in a suppression of the EISW term. 
\end{enumerate}

Going from Planck$^{\rm TT}$ to Planck$^{\rm TTTEEE}$ datasets (i.e. switching between Fig.~\ref{fig:ConstraintsCMBOnly} and Fig.~\ref{fig:ConstraintsCMBOnly2} from blue curves and contours to orange/red ones) does not alter the picture very much, but helps tightening further the allowed decaying DM fraction. In the long-lived regime, the product $\fdcdm\Gammadcdm$ is now constrained to $\fdcdm\Gammadcdm < 6.3\times10^{-3}$ Gyr$^{-1}$, which is an improvement of 7\%. In the intermediate lifetime case, one can see a ${\cal O}$(10\%) improvement: the fraction of dcdm has to be as small as 3.8\% for $\Gammadcdm \gtrsim 0.3$ Gyr$^{-1}$. This results is also in agreement with Ref.~\cite{Chudaykin:2016yfk}, which found a bound of 4\% on $\fdcdm$ in this regime for Planck TT,TE,EE + low-P but no lensing likelihood.
Finally, in the short-lived regime the constraint tightens more, by about $\mathcal{O}(20-30\%)$, or even up to a factor 2 below $\Gamma\simeq10^4$ Gyr$^{-1}$. This was expected from the previous discussion, since the impact of a varying $\Omegadm$ is strong on both EE and TE spectra, either through lensing or reionisation effects.

\subsection{Adding low redshift astronomical data}
Let us now add data from BAO, $H_0$ and matter power spectrum measurements to our dataset, in order to tighten the diagnostic power on decaying DM models. This turns out not to be very straightforward, for at least a couple of reasons:
i) a technical one is that some of these ``low redshift'' data, notably the CFHT ones \cite{Heymans:2013fya}, provide weak lensing measurements of the matter power spectrum up to $k\simeq 5$ h/Mpc.
Unfortunately the calculation of the decaying DM model matter power spectrum in the non-linear regime  raises some concerns, which will be described in Sec.~\ref{sec:NL} below.
ii) At face value, the measurements of $H_0,~\sigma_8, ~\Omega_m$ from low-redshift probes seem to be in tension with CMB-inferred  
determinations. 
For instance, the reduced Hubble parameter today derived by Planck is $h = 0.6727\pm0.0066$ (Planck 2015 TT,TE,EE+lowP \cite{Planck15}). This is about 3.0$\,\sigma$ lower than the most recent value of Ref.~\cite{Riess:2016jrr}, $h = 0.7302\pm0.0179$, obtained from the Hubble Space Telescope (HST) data. 
The CMB-inferred values of $\sigma_8$ and $\Omega_{\rm m}$ also at more than 2$\,\sigma$ than the ones coming from cluster counts \cite{Ade:2015fva} or weak lensing CFHT tomographic analysis \cite{Heymans:2013fya}. 
There are two ways out to this tension:
a) Since the comparison can only be done within a given cosmological model, it may be that $\Lambda$CDM is incomplete, and the tension would be
eventually resolved if data were analyzed within  ``the true'' cosmological model. Perhaps the decaying DM hypothesis works exactly in the sense wanted.
Further considerations on this important issue will be the topic of sec.~\ref{sec:discrepancies}.
 b) Alternatively, there are perhaps underestimated systematics in one or several of the datasets used. Without entering the difficult question of what
those errors may be, in Sec.~\ref{igndiscr} we explore the implications for decaying DM constraints of combining only data which are mutually consistent.

\subsubsection{Linear vs Non-linear matter power spectrum}
\label{sec:NL}

\begin{figure}[!htb]
\centering
\includegraphics[scale=0.4]{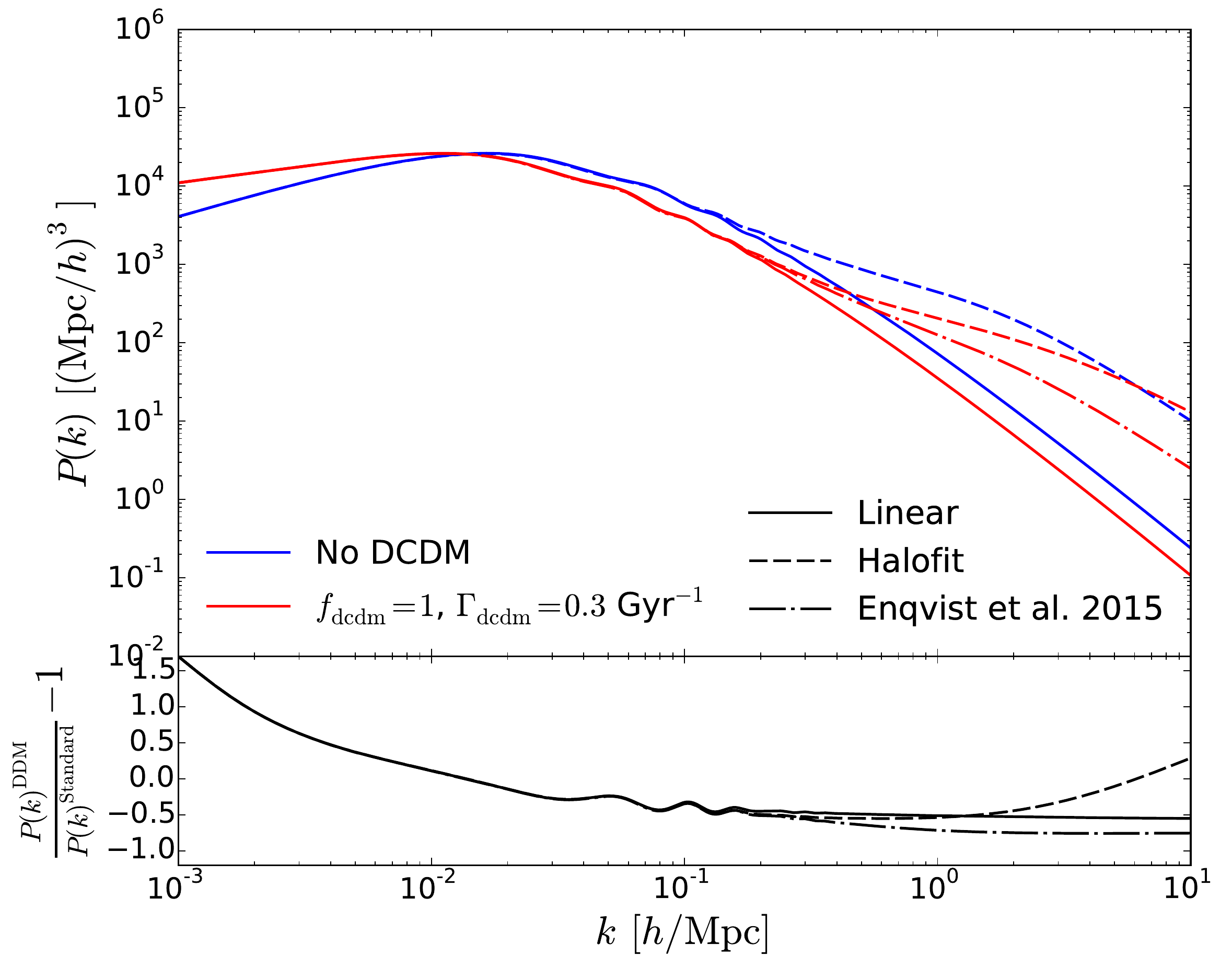}
\caption{\label{fig:NLPowerSpectrum} The non-linear matter power spectrum computed using halofit for a universe with dcdm compared to the standard $\Lambda$CDM cosmology. 
We fix all parameters to their best-fit value for Planck 2015 TT,TE, EE+low-P \cite{Planck15}: \{$\theta_s$=1.04077, $\omegadm^{\rm ini}$ or $\omega_{\rm cdm}$ = 0.1198, $\omega_{\rm b}$ = 0.02225, $\ln(10^{10}A_s)$=3.094, $n_s$=0.9645, $z_{\rm reio}$=9.9\}.}
\end{figure}
The most straightforward way to deduce the  non-linear matter power spectrum in presence of decaying DM is obviously via N-body simulations.
This approach has been considered in Ref.~\cite{Enqvist:2015ara}, in which the N-body code {\sf Gadget2} \cite{Springel:2005mi} has been modified by considering an evolving ``N-body particles'' mass. Unfortunately we cannot rely fully on their results for a couple of reasons:
first, they only consider a model in which the whole DM is decaying (i.e. $\fdcdm = 1$);  hence, by definition their results only apply to long-lived decaying DM, leaving out most of the parameter space of interest for us. The second one is perhaps more subtle: for the specific case at hand, it is unclear to which extent one can fully rely on simulations of relatively small cosmological ``boxes''. To understand why, let us remind the reader that linear theory essentially predicts that decaying DM yields a power-spectrum suppression, almost constant in fraction, at sufficiently large $k$ (see solid curve in the bottom panel of Fig.~\ref{fig:NLPowerSpectrum}). Numerical simulations of Ref.~\cite{Enqvist:2015ara} match this behaviour, eventually showing a stronger suppression at  larger $k$ (see dot-dashed curve in Fig.~\ref{fig:NLPowerSpectrum}, based on the fitting formula reported in~\cite{Enqvist:2015ara}.) However we also know that: 
i)  A strong impact of the decaying DM is to enhance the amplitude of the matter power spectrum on very large scales (very small $k$'s), as discussed in section \ref{sec:MatterPowerSpectrum}. This is again visible in the solid curve in the bottom panel of Fig.~\ref{fig:NLPowerSpectrum}, reporting the linear result.
ii) Deeply in non-linear regime, IR and UV mode-mode coupling is potentially important.  Unfortunately, the size of boxes adopted in~\cite{Enqvist:2015ara} is insufficiently large to capture this behaviour at small $k$, hence may be also missing its consequence at large $k$. 
A further caveat may be added if one compares the results of the simulations with a simple-minded use of the halofit function \cite{Smith:2002dz}, see dashed curves in Fig.~\ref{fig:NLPowerSpectrum}. Although this fitting function has been derived in the context of standard $\Lambda$CDM cosmology (extended to non-zero neutrino mass \cite{Takahashi:2012em}), the main impact of the decaying DM on the DM properties should be to slightly modify the $(1+z)^3$ dilution term, without a priori changing any of its clustering properties. One could argue that halofit is still reliable for such models, 
but unfortunately the result departs from the one of the simulations of ref.~\cite{Enqvist:2015ara}: On small scales, the non-linear power spectrum in the decaying DM universe increases and even exceeds the one in a $\Lambda$CDM cosmology. The bottom line is that it is unclear up to which values of $k$ one can rely either on the simulations of~\cite{Enqvist:2015ara} or halofit results. A safe bet is to limit the analysis to modes $k\lesssim 0.4-0.5\,h$/Mpc, where all results agree and the departure from linear theory are minor.
One implication of this cautionary approach is that we cannot rely on the full $P(k)$ from CFHT, extending to $k\simeq 5$ h/Mpc. For our considerations in the next section, we will simply make use of the inferred values of $\sigma_8\Omegam^\alpha$, which are at the hearth of the claimed discrepancies. Of course, in models departing from
$\Lambda$CDM the whole procedure used by CFHT to extract $\sigma_8\Omegam^\alpha$ should also be affected, and a different value of $\sigma_8\Omegam^\alpha$ might result. Unfortunately, we cannot check this point explicitly, and we will limit ourselves to perform analyses similar to the existing literature on this subject. 

\subsubsection{The low-redshift data discrepancies}
\label{sec:discrepancies}
\begin{figure}[!h]
\centering
\includegraphics[scale=0.5]{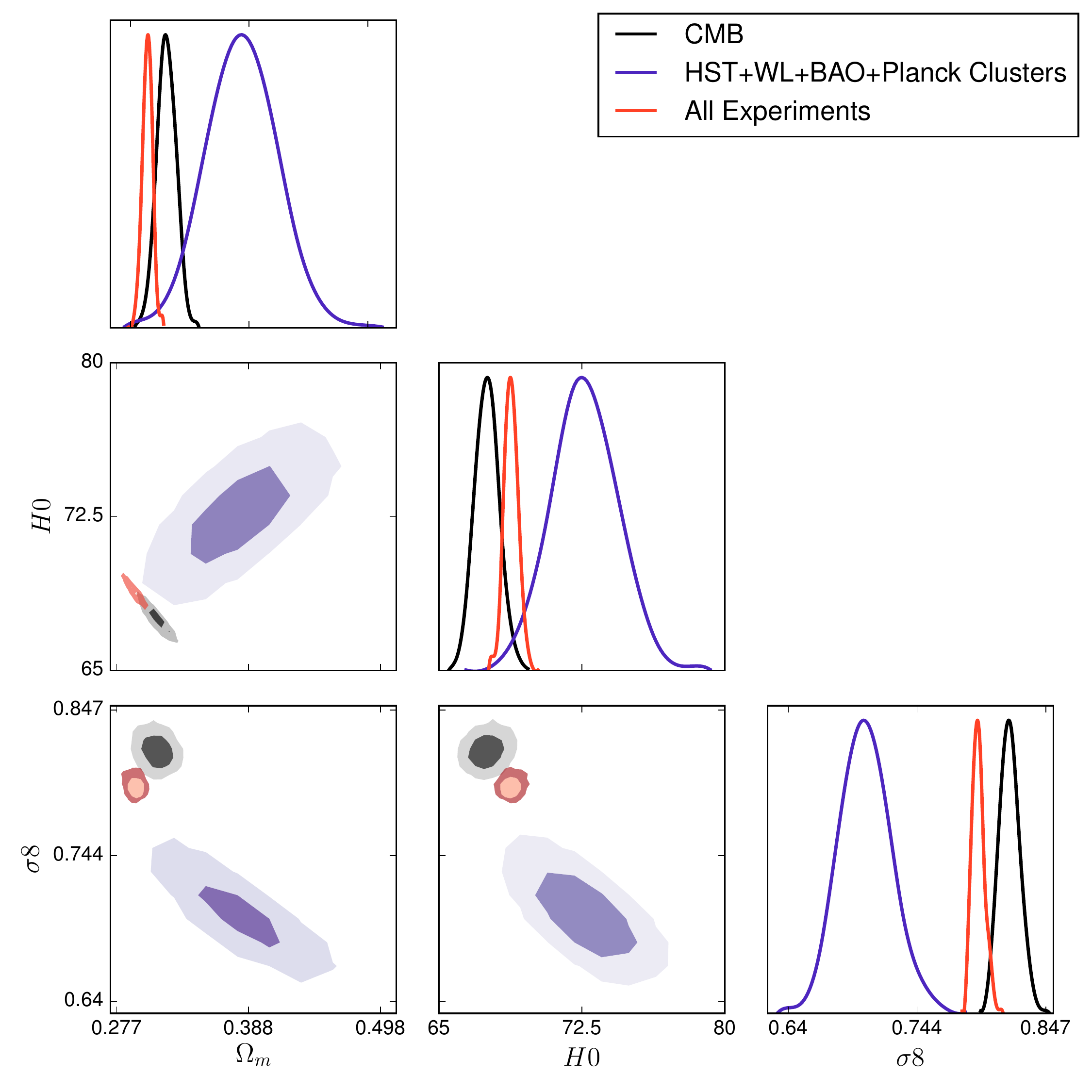}
\caption{\label{fig:7}Illustration of the discrepancies between measurements of $\sigma_8$, $\Omega_m$ and $H_0$ coming from the CMB and low-redshift experiments in the $\Lambda$CDM model. Inner and outer coloured regions denote $1\,\sigma$ and $2~\sigma$ contours, respectively.}
\end{figure}

\begin{figure}[!h]
\centering
\includegraphics[scale=0.5]{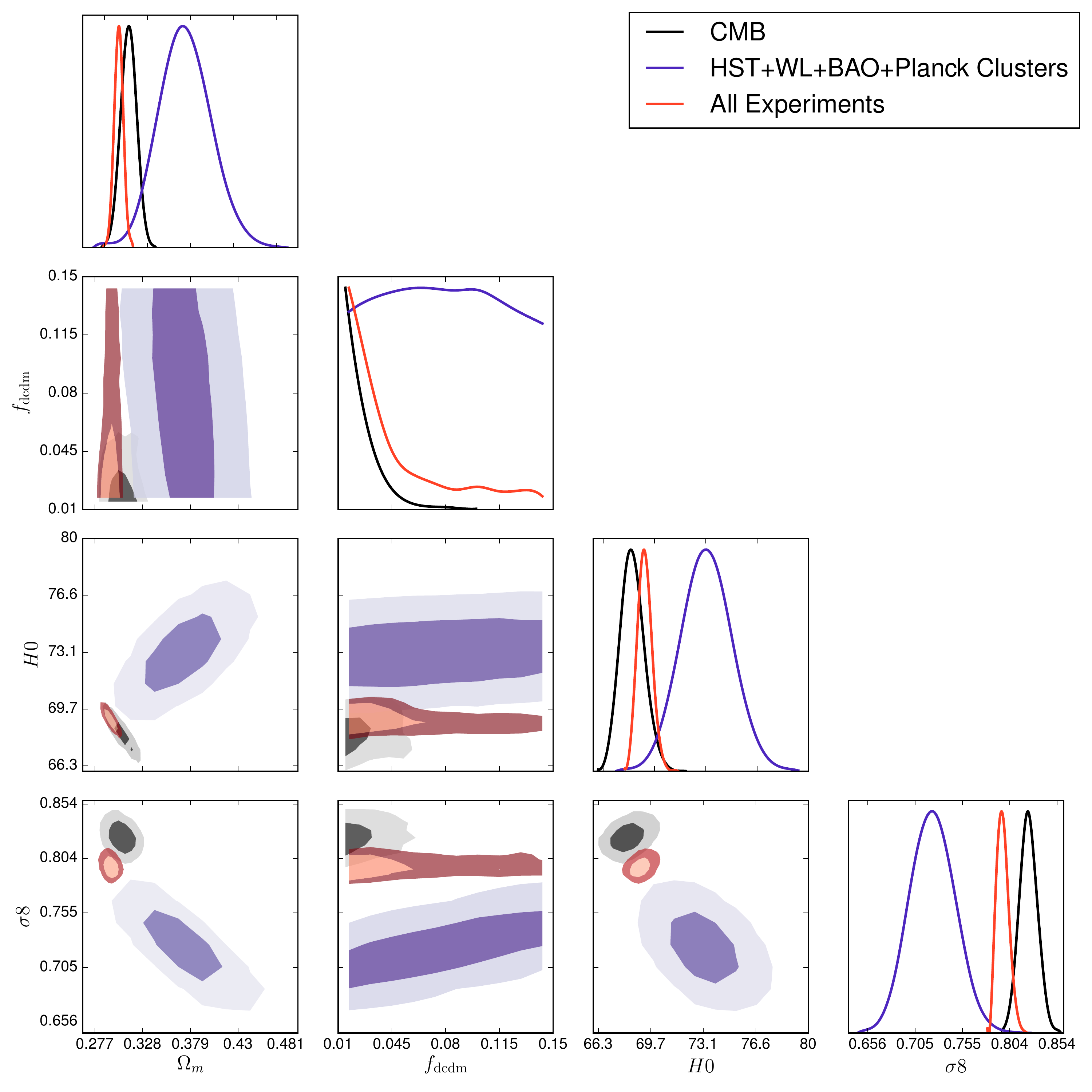}
\caption{\label{fig:8} As in Fig.~\ref{fig:7}, but in the dcdm model with $\Gammadcdm > H_0$.}
\end{figure}

\begin{figure}[!h]
\centering
\includegraphics[scale=0.5]{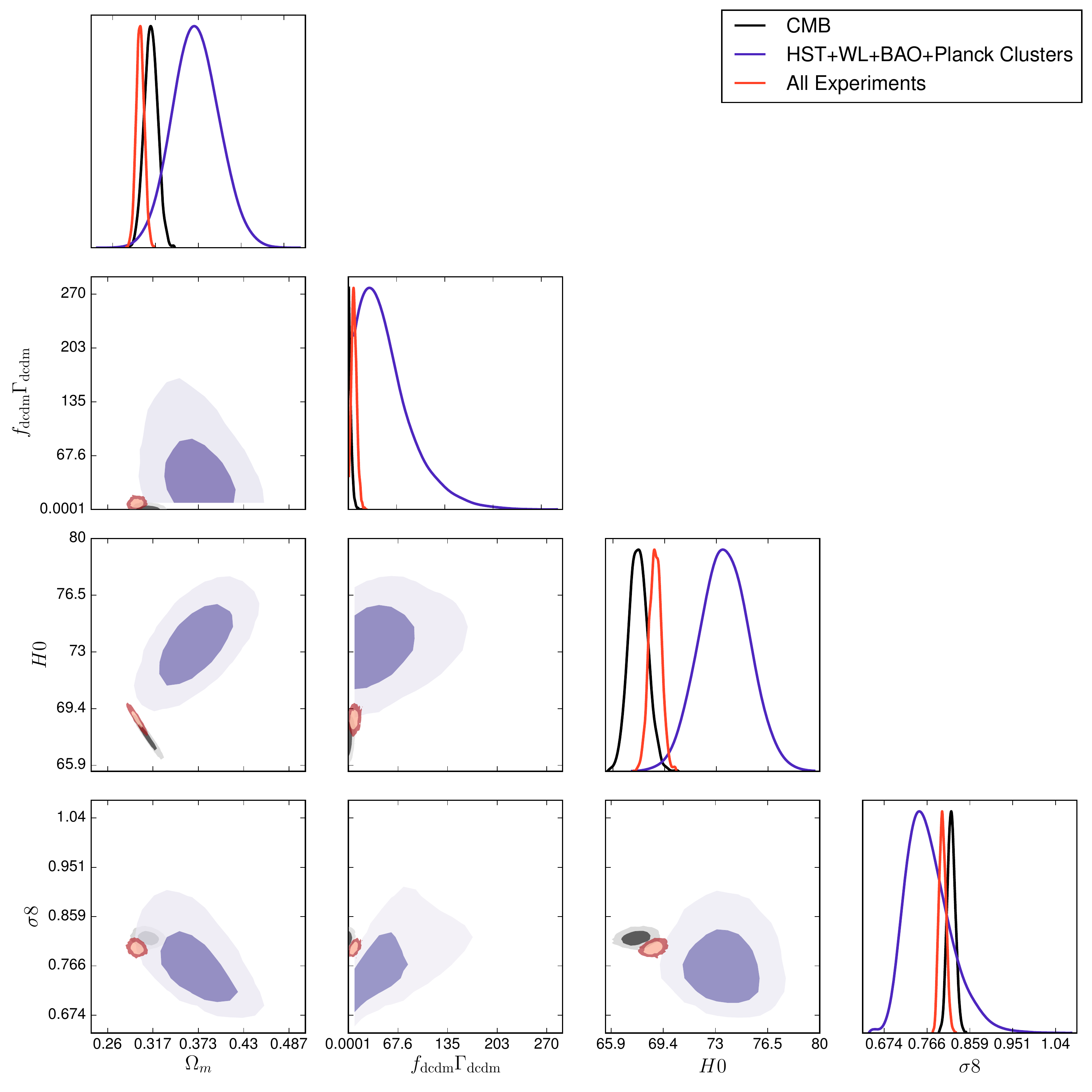}
\caption{\label{fig:9} As in Fig.~\ref{fig:7}, but in the dcdm model with $\Gammadcdm < H_0$.}
\end{figure}

In this section, let us assume that the discrepancies are physical, and thus hint to departures from the $\Lambda$CDM model.  Numerous alternative explanations are available in the literature: we can mention a possible interaction between dark matter and dark radiation~\cite{Lesgourgues:2015wza} or in the dark energy sector~\cite{Pourtsidou:2016ico}, or adding sterile neutrinos with pseudoscalar self-interactions \cite{Archidiacono:2015oma}. Actually, it has been pointed out that even decaying DM could help in solving these discrepancies \cite{Enqvist:2015ara,Berezhiani:2015yta,Chudaykin:2016yfk}.  Here we wish here to revisit these claims. 
Qualitatively, it is simple to understand why decaying DM may provide the needed ingredient. Since $\Omegadcdm$ and $\Omegam$ decrease with time and CMB data pins down very precisely the value of $\omega_{\rm cdm}\equiv\Omega_{\rm cdm} h^2$, the value of $h$ has to be bigger than in $\Lambda$CDM to compensate for the decay. Similarly, cluster count and weak lensing data\footnote{In reality, CFHT is a measure of the $P(k)$ up to scales deep in the non-linear regime. As explained in section \ref{sec:NL}, we cannot reliably modelize the non-linear growth of structure in the dcdm universe and therefore only make use of the $\sigma_8\Omegam^\alpha$ measurement.} measure the combination $\sigma_8\Omegam^\alpha$. Hence, a smaller $\Omegam$ has to be accompanied by a bigger $\sigma_8$. Thus, one could in principle hope to find a model which satisfies both CMB and low redshift astronomical data. In fact, this seems to be the case according to Ref.~\cite{Berezhiani:2015yta}, where best fit models were found for $\{\fdcdm,\Gammadcdm\}\simeq \{10\%,\,1\,{\rm Gyr}^{-1}$\}.  In practice however, we have seen that CMB is still very sensitive to modification of cosmology below the redshift of recombination, so that our analysis  based on  all CMB datasets disfavours a fraction of DM decaying between recombination and today larger than about 3.8\%; it is then reasonable to anticipate that in a global analysis no fully satisfactory solution to the tension can be found, at best marginal improvements. 
Yet, in the recent article~\cite{Chudaykin:2016yfk} a fit of a decaying DM model to Planck TT, TE, EE, lensing and cluster counts, as well as to an earlier $H_0$ determination~\cite{Riess:2011yx}, has been claimed to improve over $\Lambda$CDM by almost 2.5$\,\sigma$\footnote{These authors also underline that improvements are very sensitive to small tensions between Planck's best estimation of the lensing amplitude from the TT, TE, EE spectra and from the full lensing reconstruction using 4-point correlation functions. We do not wish to enter here into these details and refer interested readers to that paper. We simply quote their best fit result to all datasets.}. On the other hand, Ref.~\cite{Enqvist:2015ara} fitted a model with $\fdcdm =1$ to Planck TT spectrum, WMAP09 polarisations, CFHT and BAO data, and did not find more than $\sim1\,\sigma$ improvement over $\Lambda$CDM. As a contribution to clarify the  situation, we repeat the analysis, combining Planck CMB data with CFHT \cite{Heymans:2013fya} and Planck cluster \cite{Ade:2015fva} constraints on $\sigma_8\Omegam^\alpha$, BAO measurements from~\cite{Anderson:2013zyy} and the $H_0$ determination from~\cite{Riess:2016jrr}, which is an update of the former 2011 result of ref.~\cite{Riess:2011yx}.
Results are summarized in Table \ref{table:chi2} and shown in Figs. ~\ref{fig:7},~\ref{fig:8},~\ref{fig:9}. These figures just illustrate how we can  reproduce, at least qualitatively, results of the past literature: Fig.~\ref{fig:7} shows the tension in $\Lambda$CDM vs. low-redshift data, Fig.~\ref{fig:8} the improvement in a short-lived dcdm cosmology and Fig.~\ref{fig:9} in a long-lived dcdm cosmology. To gauge how important the discrepancy and the improvement are, however, let us inspect more in detail the numerical entries in Table~\ref{table:chi2}.
Within $\Lambda$CDM, one can see that the addition of the data in tension, namely HST, CFHT and Planck clusters (only 3 datapoints, dubbed
 dataset Ext$_{A}$), degrades the $\chi^2_{\rm min,eff}$, defined as $-2 \log {\rm (Likelihood)}$, by 31.2 (cf. Planck$^{\rm TTTEEE}$+Ext$_{B}$ vs Planck$^{\rm TTTEEE}$+Ext$_{A}$+Ext$_{B}$ in Table \ref{table:chi2}; the dataset Ext$_{B}$ is composed of BAO data and the WiggleZ galaxy power spectrum, in agreement with Planck CMB data). When turning to the dcdm model,  we qualitatively confirm the previous claims in that an improvement is present, corresponding to a shift in the best--fit value of $h$, and a small preference for lower $\Omega_m$ and bigger $\sigma_8$ values. However,  $\chi^2_{\rm min, eff}$ improves at most by 6.7 (slightly above $2\,\sigma$) at the price of adding two new parameters to the model; had we dealt with a satisfactory physical model, we should have expected an improvement in $\chi^2_{\rm min, eff}$ of about 30.
We therefore conclude that global fits to current data  are only marginally improved when switching to a decaying DM cosmology, at the price of complexifying the model by the addition of two new parameters. However, since there is a weak preference for a non-vanishing decaying DM fraction, in  this framework the bound on $\fdcdm\Gammadcdm$ (not surprisingly) weakens to  $\fdcdm\Gammadcdm<15.9\times10^{-3}$ Gyr$^{-1}$ at 95\% CL. 

\begin{table}
\hspace{-0.5cm}
\begin{tabular}{|l||c|cc|cc|} 
 \hline 
\multicolumn{1}{|c||}{dataset} &\multicolumn{1}{|c|} {$\Lambda$CDM} & \multicolumn{2}{|c|}{DCDM ($\Gammadcdm > H_0$)}  & \multicolumn{2}{|c|}{DCDM ($\Gammadcdm < H_0$)} \\ 

&$\chi^2_{\rm min,eff}$ 
&$\chi^2_{\rm min,eff}$  & $\fdcdm$ &
$\chi^2_{\rm min,eff}$  & $\fdcdm\Gammadcdm$ (Gyr$^{-1}$)\\
\hline
Planck$^{\rm TT}$ & 11272.3  & 11272.3 &  < 4.2\% &11272.3  &  <  6.7$\times10^{-3}$\\
Planck$^{\rm TTTEEE}$ &12952.4 & 12952.4 &  < 3.8\% & 12952.2 &  < 6.3$\times10^{-3}$ \\
Ext$_{A}$  & 4.665 & 4.19& $-$ & 3.691 & < 0.14  \\  
Planck$^{\rm TTTEEE}$+Ext$_{B}$ &13775.5  & 13775.5 &  < 3-3.6\% & 13775.5 & < 5.9$\times10^{-3}$  \\ 
Planck$^{\rm TTTEEE}$+Ext$_{A}$+Ext$_{B}$  &13806.7 &13804.1   & < 3.5-4.2\% & 13800.0 &  < 15.9$\times10^{-3}$  \\ \hline 
 \end{tabular} 
 \caption{Comparaison of the $\chi^2_{\rm min,eff}$ and constraints on $\fdcdm$ or $\fdcdm\Gammadcdm$ as a function of the dataset, in the $\Lambda$CDM and the DCDM models. The dataset Ext$_{A}$ is composed of CFHT, BAO, HST and Planck Clusters, which we refer as the discrepant dataset in the text. The dataset Ext$_{B}$ is instead composed of BAO data and the WiggleZ galaxy power spectrum, all in agreement with Planck CMB data.\label{table:chi2}}
\end{table}

\subsubsection{The strongest bounds on the decaying Dark Matter fraction and lifetime from mutually consistent data}\label{igndiscr}
\begin{figure}
\centering
\includegraphics[scale=0.8]{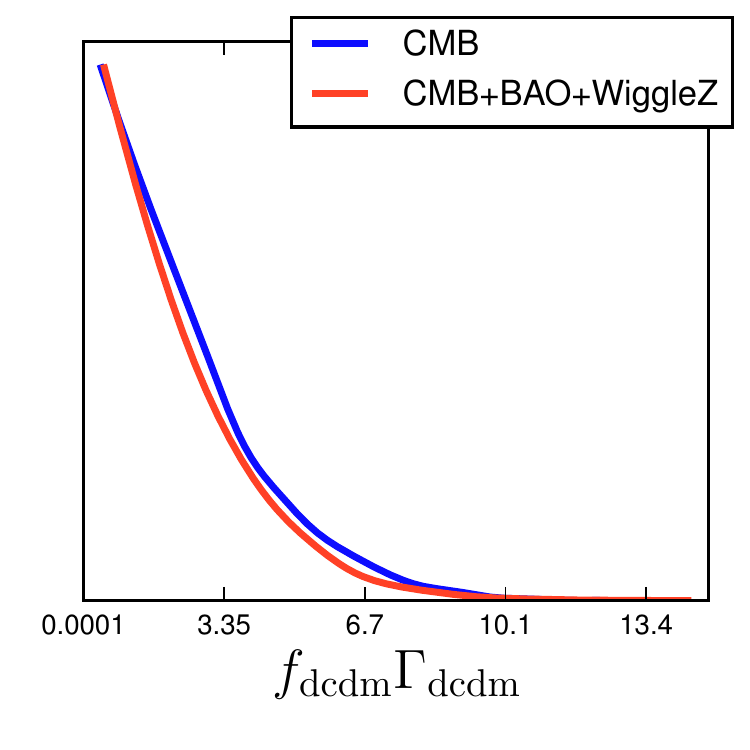}\\
\includegraphics[scale=0.6]{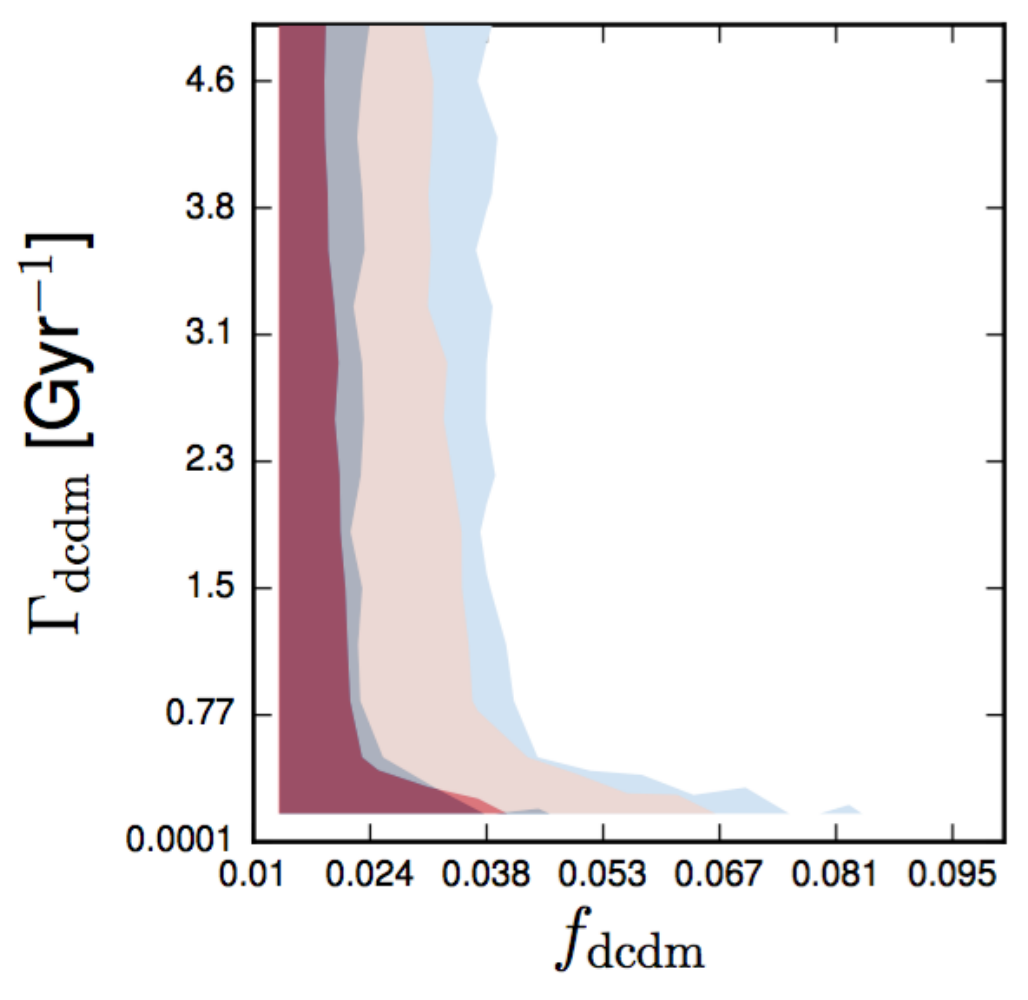}
\includegraphics[scale=0.6]{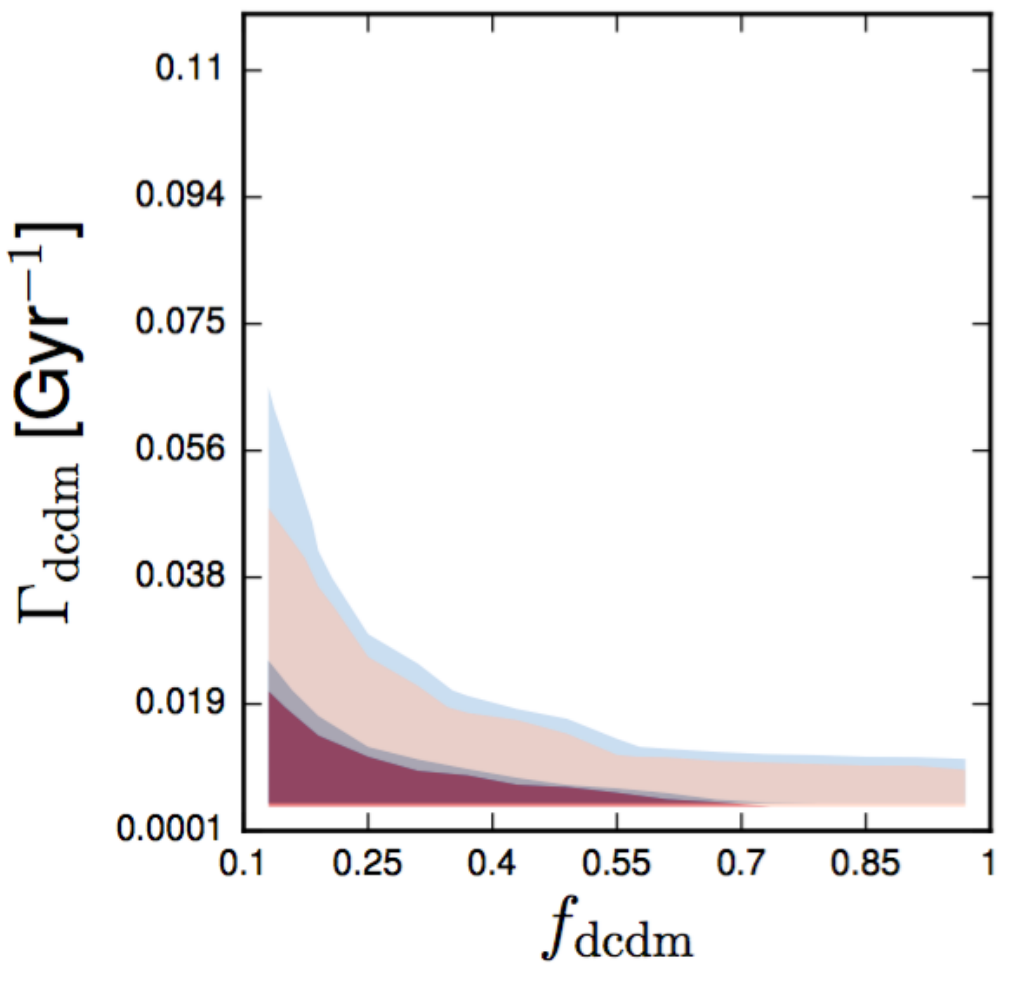}\\
\caption{\label{fig:ConstraintsAllExp1} Strongest constraints on the decaying dark matter fraction $\fdcdm$ as a function of the lifetime $\Gammadcdm$  in the long-lived and intermediate regime.} 
\end{figure}
\begin{figure}
\centering
\includegraphics[scale=0.65]{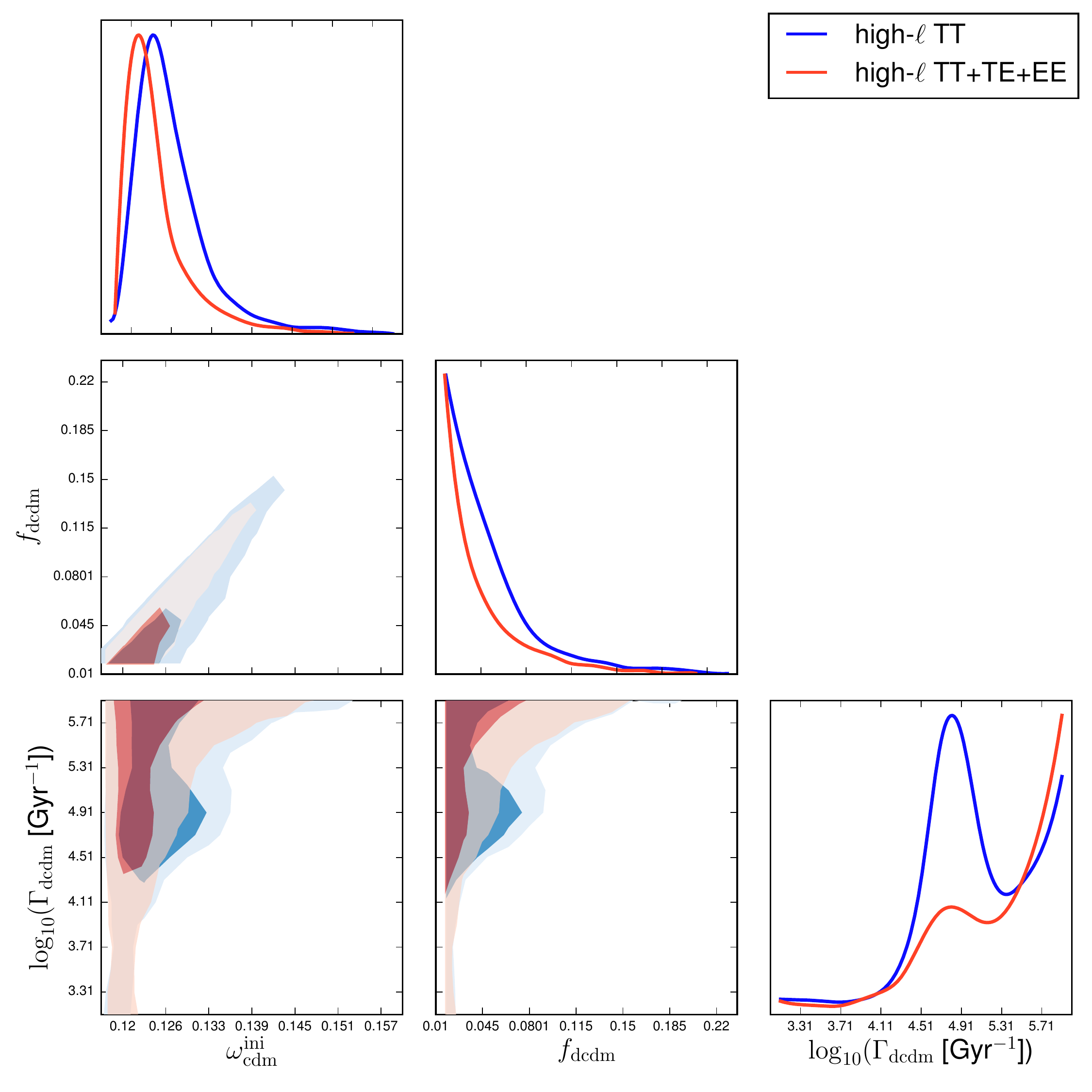}
  \caption{\label{fig:ConstraintsAllExp2} Strongest constraints on the decaying dark matter fraction $\fdcdm$ as a function of the lifetime $\Gammadcdm$ in the short-lived dcdm regime.} 
\end{figure}
Finally, if assuming that the tension between CMB and low redshift astronomical measurements are due to unknown systematics in the latter ones, one may ask what is the improvement 
on decaying DM constraints in a global analysis of {\it mutually consistent} datasets. We thus add to the Planck$^{\rm TTTEEE}$ dataset defined previously the BAO measurements at z = 0.32 and 0.57 of the BOSS collaboration~\cite{Anderson:2013zyy} and the $P(k)$ data from WiggleZ \cite{Parkinson:2012vd}\footnote{It is safe for us to use this dataset since it only probes $k\leq0.5$ h/Mpc, which is only a weakly non-linear regime, as discussed in Sec.~\ref{sec:NL}.}, collectively dubbed Ext$_{B}$. The result for the different  regimes are shown in Fig.~(\ref{fig:ConstraintsAllExp1}).

The bounds on $\fdcdm\Gammadcdm$ now tightens to $\fdcdm\Gammadcdm < 5.8\times10^{-3}$ Gyr$^{-1}$, or equivalently $\tau_{\rm dcdm}/\fdcdm>170$ Gyr. If we compare to previous result \cite{Audren14}, the use of Planck 2015 data improves only by about 6\% previous bounds derived with Planck 2013 and polarisation data from WMAP9 \cite{WMAP09}. In the intermediate regime, the additional data improve the bound, but now in a way slightly dependent on the lifetime: Roughly, the 95\% CL bound on $\fdcdm$ evolves from 3\% for $\Gammadcdm \simeq 10$ Gyr$^{-1}$ up to $3.6$\% at $\Gammadcdm \simeq 0.5$ Gyr$^{-1}$, and then relaxes as CMB-only bounds for longer lifetimes. 
Finally, in the short-lived regime constraints are improved for particles decaying around matter-radiation equality by up to a factor 2. For other lifetimes, LSS data do not tighten CMB bounds.
\section{Conclusions}
\label{sec:conclu}
In this article, we have revisited the issue of cosmological bounds on decay of a {\it fraction} of dark matter into some form of inert, or ``dark'' 
radiation, i.e. relativistic degrees of freedom not interacting electromagnetically. Within the standard model, neutrinos or gravitational waves are the only candidates with the right properties, but  beyond the standard model, additional particles may play this role. 

With respect to the past literature, we have improved in several respects: The most obvious one is that we have been using the most recent datasets available, which should ideally lead  to tighter constraints. Note also that we have corrected a mistake in the older work of \cite{Ichiki04} with a similar aim, which implies that we do not expect our constraints to match those of this reference, and does not justify a direct comparison. We have described in detail the impact of the DM decay on the TT and EE CMB power spectra, and on the matter power spectrum. This impact depends a lot on the order of magnitude of the DM lifetime. We have extended the parameter space to much smaller lifetimes, which provides a very rich phenomenology.   For the first time, we also checked that degeneracies with massive neutrinos are broken when information from the large scale structure is used. Even secondary effects like CMB lensing suffice to this purpose.
All constraints were derived using the most recent data of Planck \cite{Planck15}, BAO \cite{Anderson:2013zyy} and WiggleZ \cite{Parkinson:2012vd}.
Our results suggest that the bounds derived from 2015/16 CMB data are slightly stronger than 2013 ones derived in~\cite{Audren14}: basically CMB alone is now as constraining
as the global combination used in~\cite{Audren14}. A global analysis of mutually consistent data improves the bounds further, albeit not by much: roughly by 6\% at large lifetime,
reaching as much as a 25\% improvement at intermediate ones, before slowly degrading to the CMB-only ones for very early decays. While this points to a substantial robustness of
the cosmological bound, when adding low-redshift measurements of $\sigma_8$ \cite{Ade:2015fva,Heymans:2013fya} and $H_0$ \cite{Riess:2016jrr}, the situation is more puzzling: A tension emerges, as noted in the recent past. Although some amount of decaying DM goes
in the right direction to reconcile the discrepancy, quantitatively the situation improves only marginally: 
no evidence in favour of decaying DM can be thus inferred, but at the same time the existing tension (not surprisingly) degrades the derived bounds by a factor of $\sim 3$. It appears
more likely that the discrepancy requires either a different (and possibly major) alteration of the $\Lambda$CDM model, or derives from some yet unknown systematic effect. 

Compared to the bounds applying to the case where the totality of the DM is assumed to be unstable, the implications of our bounds on a decaying {\it fraction} of DM are much broader
and possibly far reaching. Although we do not aim at an exhaustive review of the model-dependent consequences of our results, let us just mention one interesting implication: 
Recently the possibility that a sizable if not dominant fraction of the DM is in fact in the form of stellar mass primordial black holes (BH) has been reconsidered, see for instance~\cite{Clesse:2015wea,Bird:2016dcv,Clesse:2016vqa}. Naively, this possibility is in contradiction with existing bounds (see e.g.~\cite{Ricotti:2007au}), which however could be evaded for instance if the current BH population is much more massive than the initial one, due to a rich merger history~\cite{Clesse:2016vqa}. In BH mergers, however, a sizable fraction of their mass is converted in gravitational waves. In the only merger detected to date by the LIGO detectors~\cite{Abbott:2016blz},  about 5\% of the mass was converted into gravitational radiation! Remarkably,  our CMB bound for the DM fraction converting into dark radiation has two features: i) for a large range of decay timescales, it is largely independent of the decay rate, suggesting that the bound may apply also to more complicated evolution histories than those described by a simple decay, at least to the DM fraction converted in radiation between recombination time to recent epoch. ii) numerically, if all of the DM is made of primordial BH, it excludes that in average they could have undergone even a {\it single} merger event with a fractional gravitational wave energy release comparable to the one detected by LIGO~\cite{Abbott:2016blz}.  While a specific study for a given merger history would be needed to draw strong conclusions, qualitatively this is a  new powerful argument to disfavor that a sizable fraction of DM is in the form of primordial black holes, if they undergo substantial reprocessing of the initial mass function. Actually, this is perhaps the only generic constraint that applies to primordial black hole DM candidates of {\it any mass}. 

Let us conclude with a comment: In this article, we have ignored effects associated to the recoil velocity of daughter particles in what we dubbed ``scenario 2''  in Sec.~\ref{sec:intro}. While our CMB bounds also apply to this case, typically (but this is a model-dependent statement!) more stringent cosmological and astrophysical constraints apply. This is suggested by a number of publications, such as~\cite{Aoyama:2014tga},~\cite{Cheng:2015dga}----which incidentally also conclude that in those decaying DM models it is challenging to reconcile CMB tension with  $\{\sigma_8,\,\Omega_{\rm m}\}$ determinations
from clusters---or \cite{Blackadder:2014wpa}---where an effective equation of state for the semi-relativistic daughter particle is derived and constrained with the SN Ia Hubble expansion diagram.
It is also expected that these models have a rich phenomenology at non-linear scales, possibly associated to the resolution of long-standing issues in the comparison of observed properties of 
small-scale structures with $\Lambda$CDM expectations. These studies require however dedicated simulations, which have recently started to be performed, see e.g. \cite{Wang:2014ina}.
Definitely, the study of decaying DM scenarios, notably in the non-linear clustering regime, may still reserve some surprises.
\appendix
\acknowledgments

We thank S\'ebastien Clesse for discussions. P.D.S. acknowledges support from the Alexander von Humboldt Foundation.
V.P. is supported by the ``Investissements d'avenir, Labex ENIGMASS'',  of  the  French  ANR.

\bibliographystyle{ieeetr}

\bibliography{biblio}
\end{document}